\documentclass[twocolumn,aps,pra,letterpaper,superscriptaddress, longbibliography]{revtex4-2}
\usepackage{amsmath,amssymb,amsfonts}
\usepackage{amsthm, blkarray}
\usepackage{mathtools}
\usepackage{hyperref}
\usepackage{float}
\usepackage{graphicx}
\usepackage{qcircuit}
\usepackage{enumerate}
\usepackage[dvipsnames]{xcolor}
\usepackage{physics}
\usepackage[T1]{fontenc}
\usepackage{xr}

\newcommand{\rom}[1]{\textup{\uppercase\expandafter{\romannumeral#1}}}
\newcommand{\mi}{\mathrm{i}}

\newcommand{\cC}{\mathcal{C}}

\newcommand{\cE}{\mathcal{E}}
\newcommand{\cF}{\mathcal{F}}
\newcommand{\cG}{\mathcal{G}}

\newcommand{\cI}{\mathcal{I}}

\newcommand{\cM}{\mathcal{M}}

\newcommand{\cP}{\mathcal{P}}

\newcommand{\cR}{\mathcal{R}}
\newcommand{\cS}{\mathcal{S}}

\newcommand{\cZ}{\mathcal{Z}}

\begin{document}
\title{Achieving fault tolerance against amplitude-damping noise}
\author{Akshaya Jayashankar}
\thanks{Both authors contributed equally.}
\affiliation{Department of Physics, Indian Institute of Technology Madras, Chennai, India~600036}
\author{My Duy Hoang Long}
\thanks{Both authors contributed equally.}
\affiliation{Yale-NUS College, Singapore}
\affiliation{Centre for Quantum Technologies, National University of Singapore{, Singapore}}
\author{Hui Khoon Ng}
\affiliation{Yale-NUS College, Singapore}
\affiliation{Centre for Quantum Technologies, National University of Singapore{, Singapore}}
\affiliation{MajuLab, International Joint Research Unit UMI 3654, CNRS, Universit{\'e} C{\^o}te d'Azur,
Sorbonne Universit{\'e}, National University of Singapore, Nanyang Technological University, 
Singapore}
\author{Prabha Mandayam}
\affiliation{Department of Physics, Indian Institute of Technology Madras, Chennai, India~600036}

\begin{abstract}
With the intense interest in small, noisy quantum computing devices comes the push for larger, more accurate---and hence more useful---quantum computers. While fully fault-tolerant quantum computers are, in principle, capable of achieving arbitrarily accurate calculations using devices subjected to general noise, they require immense resources far beyond our current reach. An intermediate step would be to construct quantum computers of limited accuracy enhanced by lower-level, and hence lower-cost, noise-removal techniques. This is the motivation for our work, which looks into fault-tolerant encoded quantum computation targeted at the dominant noise afflicting the quantum device. Specifically, we develop a protocol for fault-tolerant encoded quantum computing components in the presence of amplitude-damping noise, using a 4-qubit code and a recovery procedure tailored to such noise. We describe a universal set of fault-tolerant encoded gadgets and compute the pseudothreshold for the noise, below which our scheme leads to more accurate computation. Our work demonstrates the possibility of applying the ideas of quantum fault tolerance to targeted noise models, generalizing the recent pursuit of biased-noise fault tolerance beyond the usual Pauli noise models. We also illustrate how certain aspects of the standard fault tolerance intuition, largely acquired through Pauli-noise considerations, can fail in the face of more general noise.

\end{abstract}

\maketitle

\section{Introduction}

A real quantum computer is prone to noise, due to the fragile nature of the quantum states carrying the information and the unavoidable imperfections in gate operations. Scaling up to large-scale, useful quantum computers relies on the theory of quantum fault tolerance~\cite{preskill98}, a suite of methods for reliable quantum computing even with noisy memory and gates. Fault-tolerant quantum computation relies on encoding the information to be processed into physical qubits using a quantum error correcting (QEC) code. Encoded operations are performed on the qubits to manipulate the information and, in the presence of noise, these must be done in a manner that controls the spread of errors. The QEC code further allows the periodic removal of errors before they accumulate to a point where the damage is irreparable, and fault tolerance tells us how to do that even with noisy error correction operations, provided the noise is below some threshold level~\cite{knill96, aharonov96, shor_FT, gottesman_FT}. The theory of fault tolerance further includes a prescription for increasing the accuracy of quantum computation by investing more physical resources in error correction.

Starting with Shor's original proposal~\cite{shor_FT}, most fault tolerance schemes are built upon general-purpose QEC codes, such as polynomial codes~\cite{aharonov}, stabilizer codes~\cite{preskill98, gottesman_FT}, and the more recent surface codes~\cite{raus_FT}, each capable of correcting a small number of \emph{arbitrary} errors on the qubits. Fault tolerance noise thresholds have been estimated for such schemes, incorporating concatenation and recursive simulation~\cite{steane, aliferis}, magic-state distillation~\cite{magic_2005}, as well as teleportation-based approaches~\cite{knill_FT}. Current threshold estimates suggest very stringent noise control requirements of less than $10^{-4}$ probability of error per gate for the concatenated Steane code, to more relaxed ($10^{-2}$) numbers for the surface codes~\cite{fowler2009}. A similar threshold of $10^{-2}$ may also be obtained by concatenating the $[[4,2,2]]$ code with a $6$-qubit code~\cite{knill_nature}, although such a protocol requires very high resource overheads to accomplish. We refer to~\cite{cross2009} for a comparative study of the fault tolerance threshold obtained for different quantum codes, at a single level of encoding, under depolarizing noise. A more recent overview of fault-tolerant schemes using surface codes and colour codes in different dimensions may be found in~\cite{campbell_terhal2017}. 

The performance of a fault tolerance scheme depends crucially on the noise in the quantum computing device in question. The standard schemes were designed assuming no knowledge of the noise in the physical qubits---hence the reliance on codes that can deal with arbitrary errors---but, threshold estimates and how well those schemes can support accurate quantum computation, give varying perspectives depending on the underlying noise models. For example, the surface code threshold numbers are usually computed for depolarizing or at best Pauli noise on the qubits; Steane-code schemes can have more relaxed threshold numbers if one assumes depolarizing noise \cite{aliferis}, rather than the adversarial noise model used in the main analysis of Ref.~\cite{aliferis}.

This invites the question of whether one can devise fault tolerance schemes specifically tailored to the predominant noise affecting the qubits. In the current noisy intermediate-scale quantum---or NISQ~\cite{preskill_nisq}---era where getting the errors in the quantum device under control is key to progress, experimenters usually attempt to acquire knowledge of the dominant noise afflicting their quantum system, and one might expect that this knowledge can be employed usefully in the fault tolerance design, to lower the resource overheads seen in general-noise schemes, and for less stringent threshold numbers. This is borne out by the fault tolerance scheme developed in a biased noise scenario, where dephasing noise is known to be dominant~\cite{aliferis_biased, FT_assymetry}. This prescription was used to obtain a universal scheme for pulsed operations on flux qubits~\cite{aliferis_biasedExp}, taking advantage of the high degree of dephasing noise in the \textsc{cz} gate, leading to a numerical threshold estimate of $0.5\%$ for the error rate per gate operation. 

Such approaches tailored to dominant noise processes can serve as the initial steps in scaling up the quantum computer. They weaken the effect of the dominant noise on the qubits until other, originally less important, noise sources become comparable in strength and one can revert to the use of the more expensive but general-purpose fault tolerance protocols. Recent efforts along these lines include fault-tolerant constructions using surface codes tailored to dephasing noise~\cite{biasedNoise_2018, biasedNoise_2020}, the proposal to use surface codes concatenated with bosonic codes to achieve fault tolerance against photonic losses~\cite{FT_biasedGKP2020}, the design of noise-adapted codes for dominating dephasing errors in spin qubits~\cite{layden2020efficient, gross2021hardware}, and the demonstration of hardware-adapted fault-tolerant error detection in superconducting qubits~\cite{rosenblum2018fault}. 

These past examples of fault tolerance schemes for biased noise have focused on asymmetric Pauli noise --- understandably so, as standard fault tolerance theory relies heavily on classifying the effect of general noise into Pauli errors --- and dealing with those errors using general-purpose Pauli-based QEC codes. In this work, we generalise the idea of biased-noise fault tolerance to noise models and noise-adapted codes that do not make use of this Pauli-error link. In particular, we deal with amplitude-damping noise, for which noise-adapted codes of a rather different nature than Pauli-based QEC codes have been developed. Such noise-adapted codes \cite{leung,fletcherthesis,cartanform_2020} are known to offer a similar level of protection as general-purpose Pauli-based codes, when the underlying noise is amplitude-damping in nature, while using fewer physical qubits to encode each qubit of information. Amplitude damping, arising from physical processes like spontaneous decay, is a significant source of noise in many experimental quantum computing platforms \cite{scqubit}. Our work demonstrates the possibility of biased-noise fault tolerance for such noise beyond Pauli noise, and importantly, points out the failure of traditional fault tolerance intuition built from looking only at Pauli noise.

Before venturing into a detailed discussion of our work, we first summarize, in the next section (Sec.~\ref{sec:summary}), our main contributions and highlight some of the key ideas that emerge. The rest of the paper is then organised as follows. In Sec.~\ref{sec:prelim} we discuss the noise model considered here and briefly review the error-correcting properties of the $4$-qubit code. In Sec.~\ref{sec:enc_gadget} we present the basic encoded gadgets that make up our fault tolerance scheme, including the error correction gadget (Sec.~\ref{sec:ec_unit}), the logical $X$ and $Z$ gadgets (Sec.~\ref{sec:logicalX}) and the \textsc{cz} gadget (Sec.~\ref{sec:cphase}). In Sec.~\ref{sec:universal}, we show how these basic encoded gadgets can be combined to obtain a fault-tolerant universal gate set via gate-teleportation. Finally, we discuss the pseudothreshold calculation in Sec.~\ref{sec:threshold} and future directions in Sec.~\ref{sec:concl}. For better readability, many of the technical details---necessary for the full logic of our work but unimportant for the general discussion here---have been relegated to the Supplemental Material (SM) accompanying this article. Those additional points are referred to at the appropriate places in this article.

\section{Summary of contributions}\label{sec:summary}
	We develop fault-tolerant \emph{gadgets} --- composite circuits made of elementary physical operations, that achieve specific functionalities --- for qubits subjected to amplitude-damping noise. We build our fault tolerance scheme using the well-known $4$-qubit code~\cite{leung} tailor-made to deal with amplitude-damping noise. 
	
At the heart of our scheme is a fault-tolerant error correction gadget that ensures proper correction in all our logical operations. In addition, we demonstrate fault-tolerant constructions of Bell-state preparation, logical $X$ and $Z$ measurements, logical $X$ and $Z$ operations, as well as the logical controlled-$Z$ (\textsc{cz}) operation. These basic fault-tolerant gadgets are used to build a universal set of logical gadgets comprising the logical two-qubit $\textsc{cz}$ gate and the single-qubit $H$ (Hadamard), $S$ (phase), and $T$ ($\pi/8$) gates using the idea of gate teleportation. From these gadget constructions, we analytically estimate the error thresholds (or pseudothresholds) for storage and for computation, using a single layer of 4-qubit-code encoding, below which error correction provides genuine improvements in storage and computational accuracies.
	
Unlike past fault-tolerant error correction units developed for Pauli noise models that require only a Pauli ``frame change"---a classical operation requiring no quantum circuits---for the recovery, the nature of the amplitude-damping noise dictates a genuine quantum recovery for the 4-qubit code. The price to pay for using a more efficient noise-adapted, non-Pauli-type code may hence be a more complicated error correction quantum circuit. Furthermore, amplitude-damping noise leads to two types of errors that one has to correct: a damping error akin to a population decay, and an error we refer to as the off-diagonal error arising from the trace-preserving requirement of the quantum noise. Our error correction gadget thus comprises a syndrome extraction unit to first detect the two kinds of errors independently and then uses that information to apply appropriate recovery circuits.
	
Crucial to the construction of our fault-tolerant gadgets is the idea of \emph{noise-structure-preserving gates}, generalizing the idea of bias-preserving gates of the recent cat-codes discussion \cite{biaspreserving_2020, xu2021} beyond Pauli noise. That we use a code that specifically corrects amplitude-damping noise means that our gadget construction must preserve that structure, rather than generating errors uncorrectable by the 4-qubit code, after propagation through the gadget. A surprising manifestation of this requirement is the fact that, contrary to conventional fault tolerance wisdom from looking at Pauli-type noise, transversality in a logical gate construction does not guarantee a fault-tolerant gadget unless the physical gates themselves preserve the noise structure. An example is the logical controlled-NOT (\textsc{cnot}) gate. The 4-qubit code admits a transversal logical \textsc{cnot} gate. However, it cannot be made fault tolerant as a physical \textsc{cnot} gate propagates the damping error into errors not correctable by the 4-qubit code. In contrast, the physical \textsc{cz} gate is noise-structure-preserving, and hence the transversal logical \textsc{cz} is automatically fault tolerant and forms our basic two-qubit gate. Similarly, a physical \textsc{ccz} gate also adds to the set of noise-preserving gates, thereby admitting a fault-tolerant and transversal construction for the logical \textsc{ccz} gate.
	
Finally, we note that our fault tolerance scheme leads to a non-trivial pseudothreshold estimate for amplitude-damping noise, thus marking an important first step in showing that channel-adapted error correcting protocols can indeed be made fault tolerant against the dominant noise that the protocols are designed for.

\section{Preliminaries}\label{sec:prelim} 

We follow the basic framework of quantum fault tolerance developed by Aliferis \emph{et al.}~\cite{aliferis}, briefly reviewed here for completeness. In a fault-tolerant quantum computation, ideal operations are simulated by performing encoded operations on logical qubits. Encoded operations, in turn, are implemented by composite objects called gadgets which are made of elementary physical operations such as single- and two-qubit gates (including identity gates for wait times), state preparation, and measurements. We assume that the noise acts on the qubits individually, except when two qubits are participating in the same two-qubit gate. A \emph{location} refers to any one of these elementary physical operations. A location in a gadget is said to be \emph{faulty} whenever it deviates from the ideal operation, and can result in errors in the qubits storing the computational data. The key challenge is to design the gadgets in such a way as to minimize the propagation of errors due to the faults within the same encoded block. Standard fault tolerance properties, ones that our constructed gadgets must satisfy, are given in SM Sec.~A. In short, they express the ability of the fault-tolerant gadgets to give a correct (or at least correctable) output even in the presence of faults within the physical operations implementing the computation as well as the error correction.

In what follows, we refer to the elementary physical operations as \emph{unencoded} operations. Our goal is to construct the \emph{encoded} or logical gadgets corresponding to the $4$-qubit code, which are resilient against faults arising from a specific noise model, namely, the amplitude-damping channel defined in Eq.~\eqref{eq:ampdamp} below. In our scheme, we use the following unencoded operations to build the fault-tolerant encoded gadgets:  
\begin{equation} \label{eq:fault-tol}
 \{ \cP_{|+\rangle}, \cP_{|0\rangle}\} \cup \{ \cM_{X},\cM_{Z},\textsc{cnot},\textsc{cz},X, Z, S, T\} .
\end{equation}
Here,  $\cP_{|+\rangle}$ and $\cP_{|0\rangle}$  refer to the preparation of eigenstates of single-qubit $X$ and $Z$ Pauli operators, respectively, and $\cM_{X}$ and  $\cM_{Z}$ refer to measurements in the $X$ and $Z$ basis, respectively. \textsc{cnot} refers to the two-qubit controlled-\textsc{not} gate, \textsc{cz} refers to the two-qubit controlled-\textsc{Z} gate, and $X,Z, S$, and $T$ are the standard single-qubit Pauli $X$, Pauli $Z$, phase gate, and $\pi/8$ gate. Note that $|0\rangle$ is the fixed state of the amplitude-damping channel defined in Eq.~\eqref{eq:ampdamp} and is therefore inherently noiseless. We assume that rest of the gates and measurements in Eq.~\eqref{eq:fault-tol} are susceptible to noise, as described below.


Our fault tolerance construction is based on the assumption that the dominant noise process affecting the quantum device is amplitude-damping noise on each physical qubit. Amplitude damping is a simple model for describing processes like spontaneous decay from the excited state in an atomic qubit, and is a common noise source in many current quantum devices. It is described by the single-qubit completely positive (CP) and trace-preserving (TP) channel, $\cE_\mathrm{AD}(\,\cdot\,)=E_0(\,\cdot\,)E_0^\dagger + E_1(\,\cdot\,)E_1^\dagger$, with $E_0$ and $E_1$, the Kraus operators, defined as
 \begin{align}
E_0&\equiv\frac{1}{2}{\left[(1+\!\sqrt{1-p})I + (1-\!\sqrt{1-p}) Z\right]}\nonumber\\
&=|0\rangle\langle 0|+\sqrt{1-p}|1\rangle\langle 1|, \nonumber \\
\textrm{and}\quad E_1 &\equiv \frac{1}{2}\sqrt{p}(X+ \mi Y)=\sqrt p |0\rangle\langle 1| \equiv \sqrt{p}E .
\label{eq:ampdamp}
 \end{align}
Here $I$ is the qubit identity, $X, Y$, and $Z$ are the usual Pauli operators, and $|0\rangle$ and $|1\rangle$ are the eigenbasis of $Z$. $p\in[0,1]$ is the damping parameter, assumed to be a small number---corresponding to weak noise---in any setting useful for quantum computing. We denote the operator $|0\rangle\langle 1|$ simply as $E$.

In our analysis below, it will be important to separate out the different error terms in $\cE_{\mathrm{AD}}$ according to their weight in orders of $p$. To that end, we write $\cE_{\mathrm{AD}}$ as
\begin{align}\label{eq:ampdamp2}
\cE_\mathrm{AD}(\,\cdot\,)&=\tfrac{1}{4}{\left(1+\sqrt{1-p}\right)}^2\cI(\,\cdot\,)+p\cF(\,\cdot\,)\\
&\quad +{\left[\tfrac{1}{16}p^2+O(p^3)\right]}Z(\,\cdot\,)Z.\nonumber
\end{align}
where $\cI(\cdot)\equiv (\cdot)$ is the identity channel, and $\cF(\cdot)$ is the TP (but not CP) channel,
\begin{align}\label{eq:amp_fault}
\cF(\,\cdot\,)&\equiv\tfrac{1}{4}{\left[(\,\cdot\,)Z+Z(\,\cdot\,)\right]}+E(\,\cdot\,)E^{\dagger}\\ &\equiv \tfrac{1}{2}\cF_z(\,\cdot\,) + \cF_a(\,\cdot\,).\nonumber
\end{align}
We refer to $\cF$ as an \emph{error} when it affects individual qubits and refer to it as a \emph{fault} when it occurs at a certain location in a circuit. Note that a single error or fault $\cF$ can cause two different kinds of errors in the computational data carried by the qubits, since $\cF$ is a sum of two terms (i) $\cF_z = \tfrac{1}{2}{\left[(\,\cdot\,)Z+Z(\,\cdot\,)\right]}$, and (ii) $\cF_a = E(\,\cdot\,)E^{\dagger}$. Written in this manner, and neglecting the $p^2$ and higher-order terms, $\cE_{\mathrm{AD}}$ can be thought of as leading to no fault (and hence no error) when the $\cI$ part occurs, and a single fault (and hence possibly errors on the data) when the $\cF$ part occurs. 

Furthermore, we say that the qubit has an \emph{off-diagonal error} if the $\cF_z$ part remains and that the qubit has a \emph{damping error} if the $\cF_a$ remains. We may remark here that the off-diagonal error arising from amplitude-damping noise has been noted in the context of superconducting qubits and is often referred to as the \emph{backaction} error in the literature~\cite{Fz_2016, Fz_2019}. That $\cF$ is not CP means that we cannot, in principle, regard the two terms in Eq.~\ref{eq:amp_fault} as happening in some probabilistic combination.

In our setting, we assume that storage errors, gate errors, as well as the measurement errors are all due to $\cE_\mathrm{AD}$. Specifically, a noisy physical gate $\cG$ is modeled by the ideal gate followed by the noise $\cE_{AD}$ on each qubit. In the case of two-qubit gates such as the \textsc{cnot} and \textsc{cz}, we assume that a noisy gate implies an ideal gate followed by amplitude-damping noise acting on both the control and the target qubits, that is, as the joint channel $\cE_\mathrm{AD}\otimes \cE_\mathrm{AD}$ on the two qubits. A noisy measurement is modeled as an ideal measurement preceded by the noise $\cE_\mathrm{AD}$, while a noisy preparation is an ideal preparation followed by $\cE_\mathrm{AD}$. Note that the noise acts on each physical qubit individually, and is assumed to be time- and gate-independent. One could more generally regard the parameter $p$ as an upper bound on the level of amplitude damping over time and gate variations.

In a practical scenario, amplitude damping may be the main noise mechanism for idling qubits, but often not for the gate and measurement operations. One could view our proposal as the lowest-level error correction protocol in a memory device where amplitude damping is dominant for the idling period. This can be coupled with a higher-level fault-tolerance scheme capable of correcting arbitrary errors, including those that arise in the gates and measurements. Nevertheless, assuming amplitude-damping errors as the only kind of errors that occur here allows us to focus on our main goal of demonstrating the in-principle possibility of  constructing fault-tolerant circuits using QEC schemes that are adapted to non-Pauli noise models.

As the basis of our fault tolerance scheme, we make use of the well-known $4$-qubit code, originally introduced in \cite{leung} and studied in many subsequent papers (see, for example, \cite{fletcherpaper, hui_prabha}), tailored to deal with amplitude-damping noise using four physical qubits to encode a single qubit of information.
The code space $\cC$ is the span of
\begin{align}\label{eq:4qubit}
 |0\rangle_L &\equiv \tfrac{1}{\sqrt 2}(|0000\rangle +|1111\rangle)\nonumber\\
\textrm{and}\quad |1\rangle_L &\equiv \tfrac{1}{\sqrt 2}(|1100\rangle +|0011\rangle),
\end{align}
giving a single encoded, or logical, qubit of information. The code space can be regarded as stabilized by the 4-qubit Pauli subgroup generated by $XXXX$, $ZZII$, and $IIZZ$~\cite{fletcherpaper}. The logical $X$  and  $Z$ operators for the $4$-qubit code are identified as 
\begin{equation}\label{eq:logical} 
\overline{X} \equiv XXII; \overline{Z}\equiv ZIZI,
\end{equation}
up to multiplication by the stabilizer operators, of course.

The $4$-qubit code permits detection and removal of the error in the encoded information arising from a single amplitude-damping fault (understood here as an application of the $\cF$) in no more than one of the four qubits. The error-detection is achieved via a two-step syndrome extraction procedure, as originally noted in~\cite{fletcherpaper}.
\begin{itemize}
\item[\underline{Step 1}.] Measure $ZZII$ and $IIZZ$---parity measurements on qubits 1 \& 2 and 3 \& 4---on the four qubits forming the code block, giving two classical bits $s_1$ and $s_2$, respectively. Note that $s_1=0(1)$ if the $+1(-1)$ eigenvalue of $ZZII$ is obtained, whereas $s_2 = 0(1)$ if the $+1(-1)$ eigenvalue of $IIZZ$ is obtained. 
\item[\underline{Step 2}.] If $(s_1,s_2)=(0,0)$, we conclude that no damping error $\cF_a$ has been detected and proceed to correct the off-diagonal error $\cF_z$; if $(s_1,s_2)=(1,0)$, we conclude that there is a damping error $\cF_a$ in qubit 1 or 2, and measure $ZIII$ and $IZII$, yielding two further classical bits $u_1$ and $v_1$; if $(s_1, s_2)=(0,1)$, we measure $IIZI$ and $IIIZ$, for two classical bits $u_2$ and $v_2$. The $(s_1,s_2)=(1,1)$ outcome does not occur in the setting of interest.
\end{itemize}
From the extracted syndromes, we can diagnose what errors have occurred as summarized in Table~\ref{tab:synQEC}, assuming that amplitude-damping faults arose in no more than one of the four physical qubits. We note here that while either of $u_1$($u_2$) or $v_1$($v_2$) are enough to determine which qubit has a damping error, extracting both is necessary for fault-tolerant parity measurements, as discussed in Sec.~\ref{sec:ec_unit}. Some of the two-qubit amplitude-damping errors can also be diagnosed with the same syndrome measurement procedure, but we ignore them, as these are higher order than the order-$p$ terms of interest here.
\begin{table}[h] 
\begin{tabular}{c | c | c | c | c | c || l}
$s_1$ \ & \ $s_2$ \ & \ $u_1$ \ & \ $v_1$ \ & \ $u_2$ \ & \ $v_2$ \ & \textbf{Diagnosis}\\ 
\hline\hline
0&0&$\times$&$\times$&$\times$&$\times$& no damping error\\
\hline
1&0&0&1&$\times$&$\times$& Qubit 1 is damped \\
\hline
1&0&1&0&$\times$&$\times$& Qubit 2 is damped \\  
\hline
0&1&$\times$&$\times$&0&1& Qubit 3 is damped \\ 
\hline
0&1&$\times$&$\times$&1&0& Qubit 4 is damped \\ 
\end{tabular}
\caption{\label{tab:synQEC}Diagnosis of error that occurred from extracted syndrome bits, assuming amplitude-damping faults arose in no more than one of the four physical qubits. A $\times$ symbol in the table indicates the syndrome bit was not extracted; see main text. Combinations of syndrome bits that do not appear in the table correspond to events with more faults.}
\end{table}

To understand the syndrome measurement for the damping errors, consider an input code state $a|0_L\rangle+b|1_L\rangle$, with complex coefficients $a,b$ satisfying $|a|^2+|b|^2=1$. The damping error of the form $E(\,\cdot\,)E^\dagger$ on different qubits results in the states,
\begin{align}\label{eq:damping1}
\textrm{damping in qubit 1:}&\quad |01\rangle\otimes |\phi\rangle,\\ 
\textrm{damping in qubit 2:}&\quad |10\rangle\otimes |\phi\rangle,\nonumber\\
\textrm{damping in qubit 3:}&\quad |\phi\rangle\otimes |01\rangle,\nonumber\\ 
\textrm{damping in qubit 4:}&\quad |\phi\rangle\otimes |10\rangle,\nonumber
\end{align} 
where $|\phi\rangle$ is the two-qubit state,
\begin{equation}\label{eq:phi}
|\phi\rangle \equiv a|11\rangle+b|00\rangle,
\end{equation}
with the coefficients $a$ and $b$ carrying the stored information.

Once the error diagnosis is done, we perform recovery to bring the state back into the code space. The recovery is again a two-step process: 
\begin{itemize}
\item[\underline{Step 1}.] A damping error $\cF_a$ is detected by the parity measurements and is to be followed by a corresponding recovery unit. Since this error is of the form $E=|0\rangle\langle 1|$ $= \tfrac{1}{2}(X+\mi Y) = \tfrac{1}{2}(I+Z)X$, the recovery amounts to fixing the $X$ error first and then the $(I+Z)$ error. For the $X$ error, when the damping occurs in qubit $1$ or $2$, we do a single-qubit bit flip to obtain the $|11\rangle$ state on the first two qubits. A measurement of the stabilizer $XXXX$ is then done to fix the $(I+Z)$ error, simultaneously mapping the states $|1111\rangle$ and $|1100\rangle$ respectively to $|0_L\rangle$ and $|1_L\rangle$ if the measurement outcome is $+1$, or $Z_{1,2}\ket{0}_L$ and $Z_{1,2}\ket{1}_L$ if the measurement outcome is $-1$. A single-qubit phase flip is applied in the latter case, thereby bringing the state $|11\rangle\otimes |\phi\rangle$ to the code state $a|0\rangle_L+b|1\rangle_L$, spreading the information back into the four qubits. When the damping occurs in qubit $3$ or $4$, the same procedure applies, but with the roles of qubits 1 \& 2 and qubits 3 \& 4 swapped.

\item[\underline{Step 2}.] On the other hand, the off-diagonal error $\cF_z$ is not detected by the parity measurements [case $(s_1,s_2) = (0,0) $] and merits a separate recovery circuit. The error $\cF_z$ can be corrected by an optimal recovery that maximizes the fidelity between the recovered state and the original state (see \cite{fletcher}). However, for simplicity, we choose a $p$-independent recovery, namely, a measurement of the stabilizer $XXXX$. The effect of the measurement is to kill the off-diagonal error whenever the measurement outcome is $+1$, since
$(I+XXXX)(Z_i\rho+\rho Z_i)(I+XXXX)=0$
for an arbitrary state $\rho$ in the code space, where $Z_{i}$ denotes a single-qubit $Z$ operation on one of the four qubits.  
\end{itemize}
	
The syndrome extraction unit and the recovery procedures discussed above are not sufficient to construct a fault-tolerant error correction gadget. The latter requires additional parity checks and flag qubits, as explained in Sec.~\ref{sec:ec_unit} below. 
 
The $4$-qubit code is an \emph{approximate} code for amplitude-damping noise in the sense that there is remnant error after the syndrome measurement and recovery, even if the fault occurs only on a single physical qubit, the case the code is designed to deal with. One can phrase this in terms of violation of the standard Knill-Laflamme error correction conditions \cite{knill} (see Refs.~\cite{leung,hui_prabha}), but for our discussion here, we simply note that the $O(p^2)$ $Z$---or \emph{phase}---error terms in $\cE_\mathrm{AD}$ [see Eq.~\eqref{eq:ampdamp2}], necessary for ensuring the TP-nature of the channel, are neither detected nor corrected by the 4-qubit code, even though it is a single-qubit error. The $4$-qubit code only detects and corrects the order-$p$ error terms, namely, those in $\cF$. The remnant $O(p^2)$ uncorrected terms will have consequences on our fault tolerance threshold discussion later.

In what follows, we develop fault-tolerant gadgets resilient to faults that occur with probability $O(p)$, neglecting the higher-order dephasing and multi-qubit damping faults. We emphasize that, in the case of amplitude-damping noise, a single $O(p)$ fault $\cF$ at any location or a single $O(p)$ error $\cF$ in the state can correspond to a single damping error $\cF_{a}$, a single off-diagonal error $\cF_{z}$, or combination of both.

\section{Basic fault-tolerant gadgets}\label{sec:enc_gadget}
In this section, we introduce the basic fault-tolerant gadgets that constitute the building blocks of our scheme. How these units are combined to form the logical gadgets is explained in Sec.~\ref{sec:universal}. The fault tolerance of those logical gadgets is automatically ensured by the fault tolerance of the basic gadgets discussed here. As we will see, we will need the following as building blocks:
\begin{enumerate}
\item Preparation of the Bell state $|\beta_{00}\rangle=\frac{1}{\sqrt 2}{\left(|00\rangle+|11\rangle\right)}$;
\item Logical $X$ and $Z$ measurements;
\item Error correction (\textsc{ec}) gadget;
\item Logical $X$ and $Z$;
\item Logical controlled-$\textsc{Z}$ (\textsc{cz}). 
\end{enumerate}
The preparation and measurement gadgets can be constructed in a straightforward manner; the details are given in SM Sec.~B. Here, we focus on the construction of the \textsc{ec} gadget as well as the logical $X$, $Z$, and \textsc{cz} operations. In every case, the physical gates come from the elementary set given in Eq.~\eqref{eq:fault-tol}.

\subsection{Error correction gadget}\label{sec:ec_unit}
\begin{figure*}
	\centering
	\includegraphics[width=0.55\textwidth]{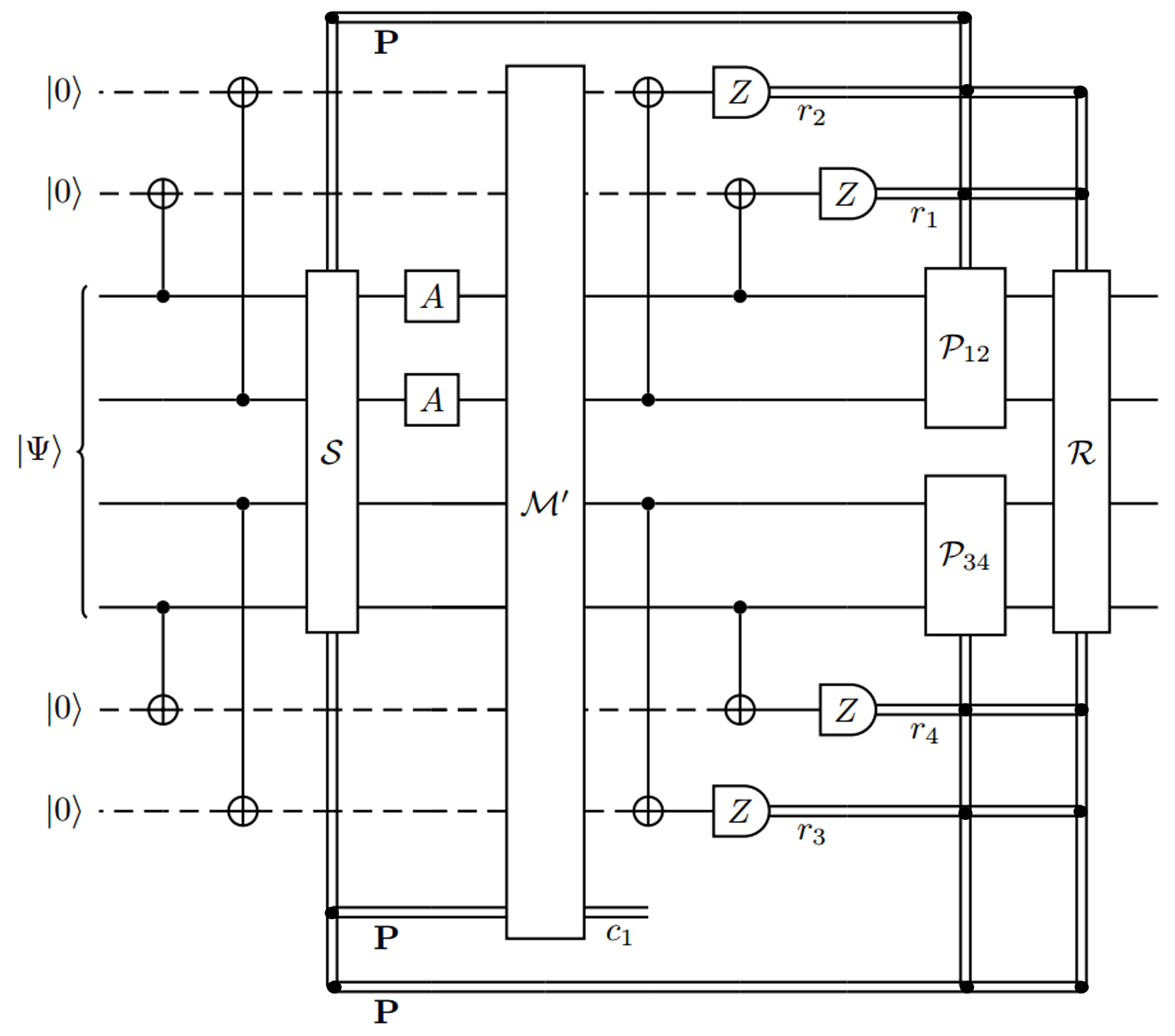}
	\caption{Circuit for the fault-tolerant \textsc{ec} gadget and the logical $X$ operation. 
	$\ket{\Psi}$ denotes the encoded data state and all the flag qubits are drawn in dashed lines. 
	$\cS$ is the syndrome extraction unit of Fig.~\ref{fig:basic_components}(d), returning the classical syndrome $\textbf{P}$. 
	$A$ is the identity gate for the EC gadget; it is the $X$ gate for the logical $X$ operation.
	If $\textbf{P}$ indicates a damping error, $\cM'$ is the identity operation; if $\textbf{P}$ indicates no damping error, $\cM'$ is the $\cM_1$ sub-unit of Fig.~\ref{fig:basic_components}(f).
	The flag qubits are then decoupled from the data qubits, with extracted classical bits $ \textbf{r}\equiv \{r_1,\ldots, r_4\}$. 
	$\cR$ is the recovery unit of Fig.~\ref{fig:basic_components}(e). If $\textbf{P}$ indicates a damping error, $\cR$ depends only on $\textbf{P}$. If $\textbf{P}$ indicates no damping error and if at least one of the two flag bits $r_1,r_2~(r_3,r_4)$ is flipped, a parity measurement $\cP$ (see Fig.~\ref{fig:basic_components}(a)) is performed on qubits 1 and 2 (3 and 4) giving syndrome bit $s_{12}$ ($s_{34}$). 
	$\cR$ is then applied, which corrects errors based on $\textbf{r}$ and $\textbf{s} \equiv \{s_{12}, s_{34}\}$. The outcome of $\cM'$ unit, $c_1$, is not necessary to correct for an error in the \textsc{ec} gadget; it is however useful in the \textsc{cz} gadget, to detect a propagated error from one block to the other (see Sec.\ref{sec:cphase}). The full details of what recovery gates to apply are given in Tab.~I and II of the SM.}
	\label{fig:ec}
\end{figure*}

\begin{figure*}
\includegraphics[width=\textwidth]{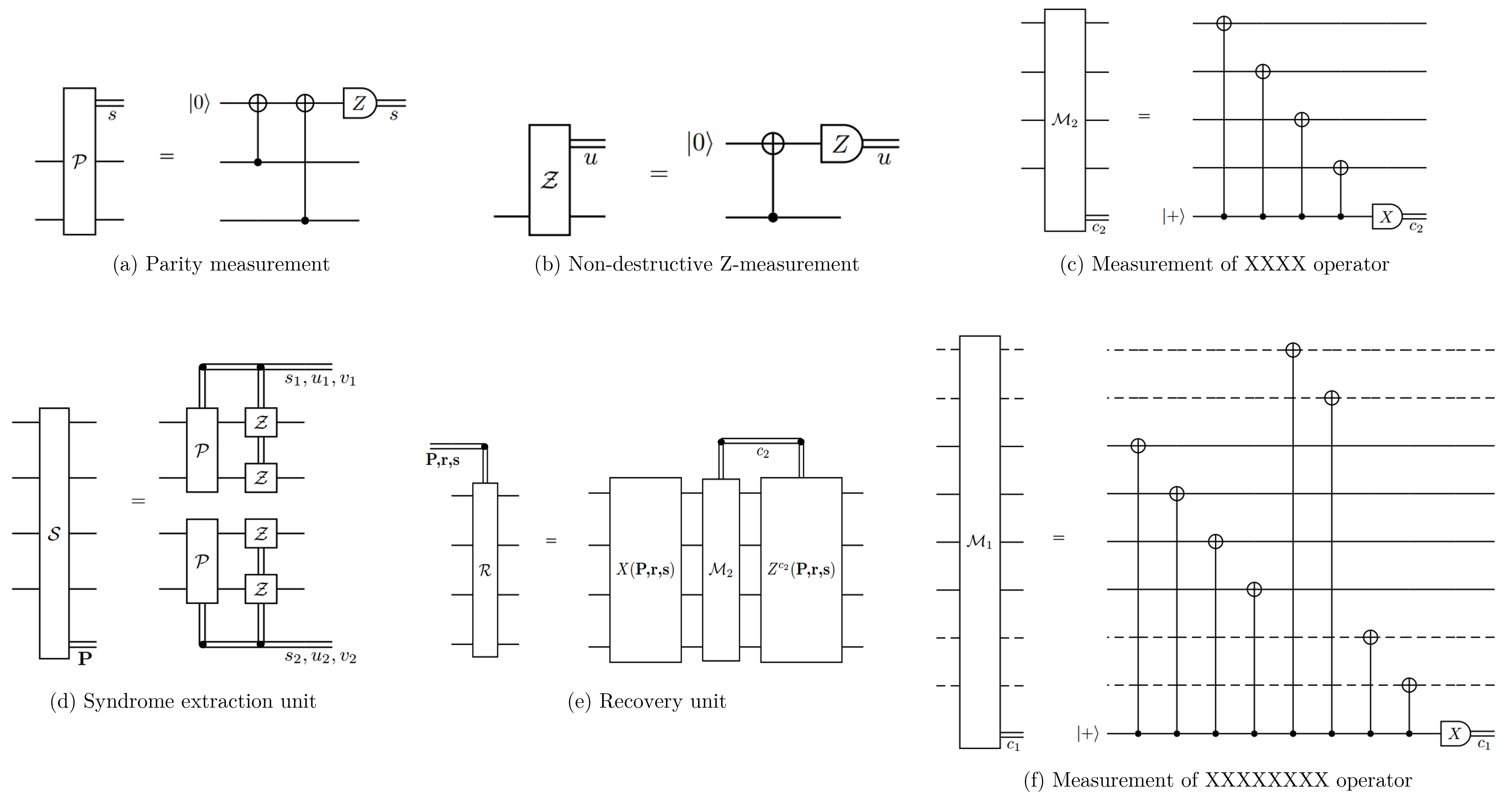}
\caption{Basic circuit components. (a) Parity measurement $\cP$ between two data qubits, giving parity bit $s$. (b) Non-destructive $Z$ measurement of a qubit, giving classical bit $u$. (c) Measurement $\cM_2$ of the $XXXX$ stabilizer on the incoming qubits, giving classical bit $c_2$. (d) The syndrome extraction unit  $\cS$ is a two-step procedure. Two parity measurements are performed on data qubits 1 \& 2 and 3 \& 4, giving syndrome bits $s_1$ and $s_2$, respectively. If $s_1=1$($s_2=1$), non-destructive $Z$ measurements are performed on qubits 1\&2(3\&4) to obtain syndrome bits $u_1, v_1$($u_2, v_2$). (e) The recovery circuit $\cR$ for damping error $\cF_a$, for incoming syndrome syndrome bits $\mathbf{P}\equiv \{s_1,s_2,u_1,v_1,u_2,v_2\}$, $\mathbf{r}\equiv \{r_1,r_2,r_3,r_4\}$, and $\textbf{s} \equiv \{s_{12}, s_{34}\}$. In $\cR$, $X$ gates are first applied, as determined by $\mathbf{P}, \mathbf{r}$, and $\mathbf{s}$. This is followed by the measurement $\cM_2$ [shown in 2(c)], followed by an application of $Z$ if $c_2=1$. }
\label{fig:basic_components}
\end{figure*}

The error correction gadget --- henceforth referred to as the \textsc{ec} gadget  --- shown in Fig.~\ref{fig:ec}, implements the syndrome extraction and the recovery procedures described in Sec.~\ref{sec:prelim}. The four qubits carrying the encoded information --- henceforth referred to as \emph{data qubits} to distinguish them from the ancillary qubits --- are in some generic state $|\Psi\rangle$. 

The circuits for a parity measurement, denoted as $\cP$, and for a non-destructive $Z$-measurement, denoted as $\cZ$, are detailed in Figs.~\ref{fig:basic_components}(a) and ~\ref{fig:basic_components}(b), respectively. Each measurement uses one ancillary qubit initialized to the $\ket{0}$ state. The circuit for the $XXXX$ measurement, denoted as $\cM_2$  and shown in Fig.~\ref{fig:basic_components}(c), uses one ancillary qubit initialized to the $\ket{+}$ state. 
	
The syndrome extraction unit $\cS$ [see Fig.~\ref{fig:basic_components}(d)] consists of two parity measurements, followed by two non-destructive $Z$-measurements to extract the position of the damped qubit, in case one of the measurement outcomes of the first two parity measurements is nontrivial. Note that a fault at the target of the first \textsc{cnot} in a parity measurement may lead to an outcome $1$ even though there is no damped data qubit. Thus, the extraction of syndrome bits from \emph{both} the data qubits in case of a non-trivial parity measurement outcome is necessary to make the syndrome extraction unit fault tolerant. 

The circuit for the recovery from a damping error, following the diagnosis of a nontrivial error, is detailed in Figs.~\ref{fig:basic_components}(e). It comprises two parts, the first part performing the bit-flip converting the data-qubit-pair with the amplitude-damping error to the state $|11\rangle$ [see Eq.~\eqref{eq:damping1}], and the second part performing a measurement $\cM_2$ of the stabilizer $XXXX$. The latter measurement, with outcome denoted as $c_2$, projects the state of the data qubits either into the code space, corresponding to the subspace with eigenvalue $+1$ ($c_2=0$), or to the subspace with eigenvalue $-1$ ($c_2=1$). The latter case corresponds to a single-qubit $Z$ error and we apply a suitable local $Z$ gate from the set $\{ZIII,IZII,IIZI,IIIZ\}$, to correct for it. For example, if the first data qubit is damped and $c_2=1$, we can apply $Z$ to the first data qubit or to the second data qubit (since $ZZII$ is a stabilizer).

In case no damping error is detected, we proceed to the recovery for the off-diagonal error $\cF_z$, which is simply the measurement of the stabilizer $XXXX$. However, this procedure is not fault tolerant due to the following reason. It is possible that the syndrome extraction unit detects no damping error ($s_1=s_2=0$), but actually there is one in the output of the syndrome extraction unit due to a faulty \textsc{cnot} at the control. If we proceed to measure the $XXXX$ operator, the damping error $\cF_a$ becomes either an $X$ or a $Y$ error, which is uncorrectable by the 4-qubit code. This can be seen by noting that, the effect of the measurement after the action of a damping error $\cF_a$ at the $k$-th data qubit -- denoted as $E_k$ -- on a state $\ket{\Psi}$ in the code space, is given by $\tfrac{1}{2}(1 \pm XXXX)E_k\ket{\Psi}=\tfrac{1}{2}(E_k \pm E_k^{\dagger})\ket{\Psi}$.

Our solution for this issue is that we just perform the $XXXX$ measurement anyway at every error correction step, but with additional \emph{flag qubits} \cite{chao2018quantum} that are added to detect faults that lead to uncorrectable errors. The circuit in Fig.~\ref{fig:ec} implements this strategy with the flag qubits marked in dashed lines and represents our fault-tolerant \textsc{ec} gadget. We conclude this section with a brief description of the inner workings of our fault-tolerant \textsc{ec} gadget and leave the detailed proofs to SM Sec.~B3. 

At the beginning of an error correction step, each data qubit is coupled to an ancillary qubit initialized in the $\ket{0}$ state, referred to as a flag qubit. We then proceed with the usual error correction procedure, starting with the syndrome extraction unit. If the syndrome extraction unit detects a nontrivial error, we decouple the flag qubits from the data qubits and use the recovery $\cR$ to correct for damping errors, based on the extracted syndrome \textbf{P}. In case there is no damping error detected, we continue with the recovery involving the $XXXX$ measurement. 

However, measuring the $XXXX$ operator on the four data qubits no longer kills off $\cF_z$ error, as it was originally supposed to do, because the four data qubits are now coupled to the four flag qubits. An $XXXX$ measurement on four data qubits alone before the decoupling step is equivalent to an $XXXXXXXX$ measurement on all data and flag qubits after the decoupling step (this can be seen, for example, by commuting the \textsc{cnot}s of the measurement step through the \textsc{cnot}s of the decoupling step). Since the flag qubits are initialized in state $\ket{0000}$, which is not a stabilized state of $XXXX$, the measurement of $XXXXXXXX$ will not kill off the off-diagonal term $\cF_z$. If the flag qubits are initialized in a stabilized state of $XXXX$, for example, $\tfrac{1}{\sqrt{2}}(\ket{0000} + \ket{1111})$, then the measurement will kill off $\cF_z$. However, the preparation of this state is also not an easy task, therefore, we instead modify the recovery by measuring $XXXXXXXX$ on all data and flag qubits before the decoupling step. This is equivalent to measuring $XXXX$ on the $4$ data qubits alone after the decoupling step, which can kill off $\cF_z$ errors. The circuit for the modified measurement, denoted as $\cM_1$, is shown in detail in Fig.~\ref{fig:basic_components}(f). We note that the circuit for $\cM_1$ is not obviously fault tolerant because a single fault at the control of one of the \textsc{cnot}s may cause multiple errors in the data qubits. However, with the use of the same set of flag qubits, and \textsc{cnot} gates performed in a certain order, this circuit can indeed be made fault tolerant (see SM Sec.~B3). 

At the end of the recovery procedure, the flag qubits are decoupled from the data qubits, and then measured in the $Z$ basis, resulting in the four bits $\{r_1,r_2,r_3,r_4\}$, denoted as \textbf{r}. If there is no fault up to this step, the measurement outcomes will be $(0000)$. Otherwise, if a data qubit is damped, the corresponding outcome of the flag qubit coupled to it will be flipped. However, notice that a fault at a flag qubit may also flip the flag outcomes, hence, we still need to distinguish between a fault in the data qubits and one in the flag qubits. To do so, we perform one more round of parity measurements, denoted as $\cP_{12}$ and $\cP_{34}$, and correct for the corresponding errors. Specifically, if the extracted syndrome $\textbf{s} \equiv \{s_{12}, s_{34}\}$ is trivial, the fault is in the flag qubits and we only need to correct for a $Z$ error using the measurement of $XXXX$ on the data qubits. Otherwise, if the syndrome is nontrivial, the fault is in the data qubits and we also need to correct for an $X$ error. The recovery unit $\cR$ is then performed to correct for the error, based on the extracted syndrome $\textbf{r}$ and $\textbf{s}$.

The \textsc{ec} gadget in Fig.~\ref{fig:ec} is fault-tolerant in the following sense: A single error $\cF$ in the incoming data-qubit state, or a single fault $\cF$ in the \textsc{EC} gadget results in no more than a single correctable (by the $4$-qubit code) error in the outgoing state of the data qubits. A detailed proof is presented in SM Sec.~B3, but the ideas can be intuitively understood as follows. If there is one damping error $\cF_a$ in the incoming state and no fault in the \textsc{ec} unit, the syndrome extraction unit will detect it and the recovery unit will correct it, as promised by the $4$-qubit code. If there is one off-diagonal error $\cF_z$ in the incoming state or in the syndrome extraction unit, it will be killed off by the $\cM_1$ unit even though it is not detected by the syndrome extraction unit. On the other hand, a single damping error $\cF_a$ in the syndrome extraction unit, in the $\cM_1$ unit, or in the flag qubits is detected by the set of four flag qubits. A fault in the ancilla used in the $\cM_1$ unit propagates $X$ errors to the data qubits which are also taken care of by the flag qubits. The outcome of the $\cM_1$ unit with $c_1=0$ would mean that the off-diagonal error $\cF_z$ on the incoming state has been killed. However, $c_1=1$ would mean that a damping error $\cF_a$ must have occurred on one of the data qubits, flag qubits or the ancilla qubit. Depending on which qubit has had the error, one could have a $ZX$ or $X$ error propagating at the output of $\cM_1$ unit. Both of these errors can be identified using the flag syndrome bits $\textbf{r}$ and corrected in the subsequent recovery unit. Since the recovery unit involves $XXXX$ measurement, the $Z$ error gets fixed without any information about the outcome $c_1$ from $\cM_1$ unit.

We note here that, unlike in standard fault tolerance analysis dealing with Pauli errors where a classical frame-change is all that is needed to correct the detected errors, here, we need a nontrivial recovery unit to correct for the single damping errors. This is due to the fact that the elementary gate operations used in our gadget constructions are not \emph{amplitude-damping preserving}: A single damping error propagates through some of the elementary gates (like $X$) into other kinds of errors, not correctable by the $4$-qubit code tailor-made for removing damping errors. Any damping error thus has to be genuinely corrected, before the next gadget can be implemented. Note that, the final local $Z$ gate, controlled by $c_2$ [i.e., $Z^{c_2}(\mathbf{P},\mathbf{r},\mathbf{s})$], in the recovery unit \emph{does} commute with subsequent damping errors and all gates in our elementary gate set and thus can, in principle, be fixed by a Pauli frame-change rather than an actual gate operation. For simplicity, however, we have kept it as a part of the recovery unit here.


\subsection{Logical $Z$ and $X$ gadgets}\label{sec:logicalX}

In standard fault tolerance schemes making use of Pauli-based codes, an operator like $\overline X=XXII$ and $\overline Z=ZIZI$ (or alternatively, $IIXX$ and $IZIZ$, with the two differing by a stabilizer operator) can be applied simply by performing $X$ or $Z$ on two of four physical qubits, the fault tolerance guaranteed by the transversal nature of the operation. In the case of the amplitude-damping code, however, only the transversal logical $Z$ is fault tolerant because the off-diagonal error $\cF_z$ and the damping error $\cF_a$ commutes and anticommutes, respectively, with a physical $Z$ gate. The transversal logical $X$ operation is no longer fault tolerant, due to the fact that the damping error of the form $\tfrac{1}{2}(X+\mi Y)$ becomes $\tfrac{1}{2}(X-\mi Y)$ after conjugating past the $X$ operator. 

Instead, to obtain a fault-tolerant logical $X$, we use the same technique as in the \textsc{ec} gadget, by making use of flag qubits. The logical $X$ gadget is given in Fig.~\ref{fig:ec}, with the same structure as the \textsc{ec} gadget. If the syndrome extraction unit detects a damping error $\cF_a$, we can apply the transversal $\overline{X}$ because the single fault allowed for the unit has already occurred. If no damping error is detected, the transversal, non-fault-tolerant $\overline{X}$ is applied first, followed by the error correction. 

The fault-tolerant properties of the logical $X$ gadget are explained in detail in SM Sec.~B4, although they mostly follow from the fault tolerance of the \textsc{ec} gadget. The main difference between this gadget and the \textsc{ec} gadget is that a damping error $E$ on a data qubit becomes $E^{\dagger}$ after conjugating through an $X$ gate. Thus an incoming error to the $\cM_1$ unit can be either $E$ or $E^{\dagger}$. However, these are single-qubit errors and the set of flag qubits is still enough to detect which qubit has the error. We also note that the flag syndromes of the logical $X$ gadget differ from that of the standard \textsc{ec} gadget by two bit flips on flag qubits 1 and 2, due to the application of two $X$ gates on data qubits 1 and 2. For example, without any faults, the flag syndrome is $(1100)$ instead of $(0000)$ as in the case of the standard $\textsc{ec}$ gadget.

\subsection{Logical \textsc{cz} gadget}\label{sec:cphase}

We next demonstrate a fault-tolerant two-qubit logical \textsc{cz} operation, an essential ingredient for realising a universal set of logical gates. We first note that the logical \textsc{cnot} and the \textsc{cz} gadgets for the $4$-qubit code both admit transversal constructions. However, as noted earlier in the construction of the $X$ gadget, transversality does not automatically translate into fault tolerance in the case of amplitude-damping errors and the $4$-qubit code. In fact, the transversal \textsc{cnot} is not fault-tolerant to amplitude-damping noise: a single error caused by the amplitude-damping noise can propagate through the transversal circuit into an error that is not correctable by the $4$-qubit code.

For example, observe that, for two physical qubits connected by a physical \textsc{cnot} operation, an incoming damping error $E$ [see Eq.~\eqref{eq:ampdamp}] on the control qubit propagates after the \textsc{cnot} into an $X$ error on the target. Meanwhile, a damping error on the target qubit propagates into $\tfrac{1}{2}(I_c X_t + Z_c Z_tX_t)$, where the subscript $c$ denotes the control qubit and $t$ denotes the target qubit. By tracing out the control qubit, we get two types of errors on the target qubit, namely, the damping error $E = \tfrac{1}{2}(1+Z)X$ and its conjugate $E^{\dagger}=\tfrac{1}{2}(1-Z)X$. We know that the $4$-qubit code cannot correct for both of these errors. A single fault on one of the qubits can thus result in an uncorrectable error, violating the requirements of fault tolerance, despite the transversal structure. 

\begin{figure}
	\centering
	\includegraphics[scale=.15]{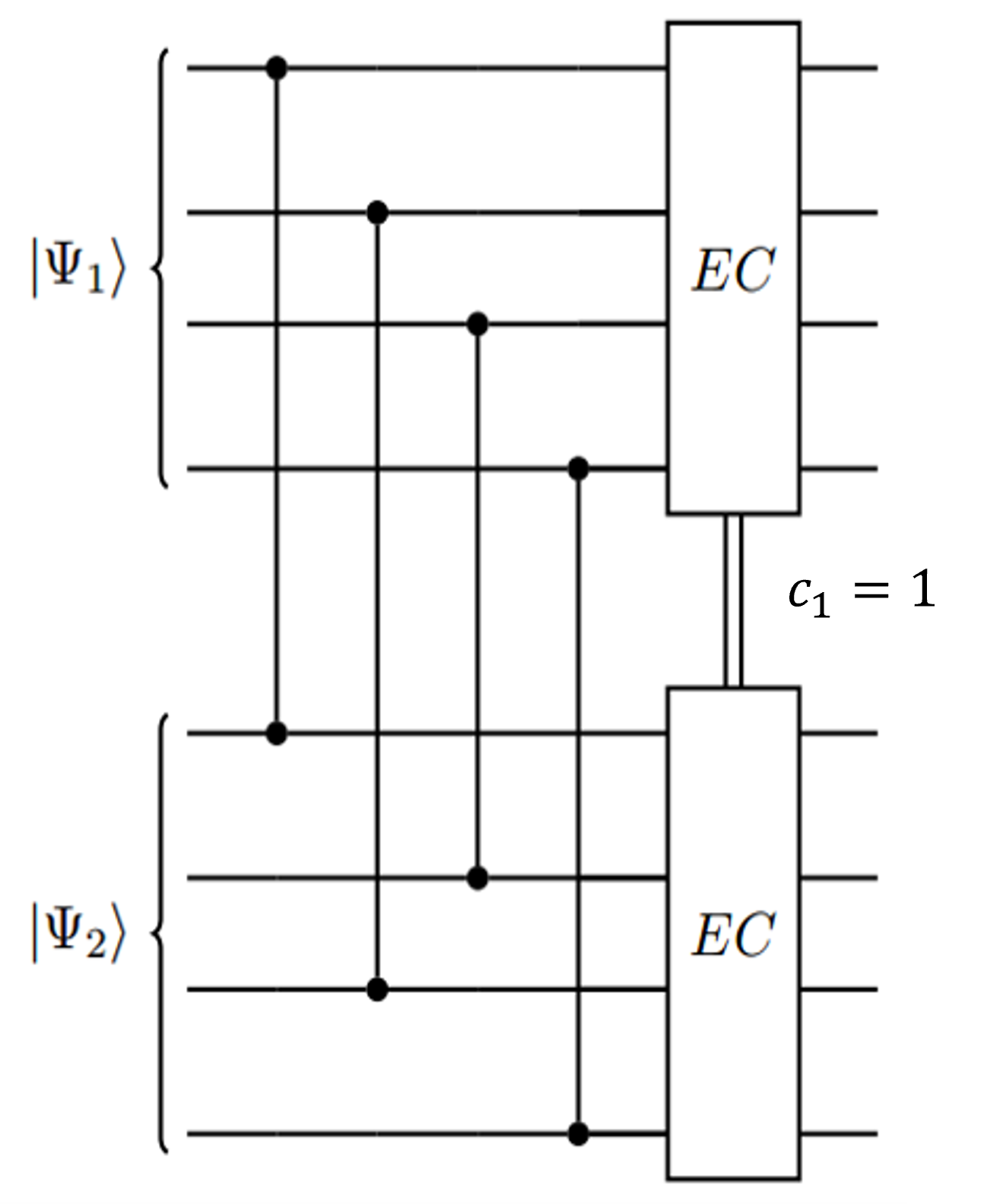}
	\caption{Fault-tolerant logical $\textsc{cz}$ gadget. $\ket{\Psi_1}$ and $\ket{\Psi_2}$ are two blocks of data qubits. The double lines between the two \textsc{ec} gadgets indicates that if one of them detects a damping error and $\cM_1$ unit in the other has outcome $+1$, then a $Z$ operator is applied to the qubit of the latter block that is connected to the damped qubit of the first block by a \textsc{cz}.}
	\label{fig:cphase}
\end{figure}  

This suggests the idea of \emph{noise-structure preserving gates}, as an important tool for fault-tolerant implementation of noise-adapted codes. Indeed, unlike the \textsc{cnot}, it turns out that the transversal \textsc{cz} gadget shown in Fig.~\ref{fig:cphase} is fault-tolerant against amplitude-damping noise. This is explained in detail in SM Sec.~B5. The basic idea, however, is easy to understand by contrasting with the \textsc{cnot} gadget: a damping error at the control (target), after propagating through a physical \textsc{cz} gate, propagates as a damping error at the control (target). However, the damping error at the control (target) of the \textsc{cz}, does lead to an additional phase ($Z$) error in the target (control).  This explains the dependence between two trailing \textsc{ec}s in the \textsc{cz} gadget, indicated by the double lines in Fig.~\ref{fig:cphase}. Whenever one of the two \textsc{ec} gadgets detects a damping error in the incoming state, and the $\cM_1$ unit in the other unit has outcome $+1$, a local $Z$ recovery operator is applied on the qubit in the latter block corresponding to the damped qubit in the first \textsc{ec} gadget. For example, if the syndrome extraction unit in the first data block detects a damping error at the second qubit and the $\cM_1$ unit in the \textsc{ec} of the second data block has outcome $+1$, a $Z$ operator will be applied to the third data qubit of the second data block, since the two qubits are connected by a \textsc{cz} gate.

Furthermore, we note that the logical \textsc{ccz} gate can also be constructed in a fault-tolerant way using sets of transversal \textsc{ccz} physical gates, executed in two time-steps (see SM Sec.~B6 for the circuit). Similar to a \textsc{cz} physical gate, a \textsc{ccz} physical gate also propagates phase error(s) to the other two qubits whenever there is an incoming damping error to one of the controls. However, a faulty control (target) itself does not propagate any error to the other two qubits. Errors generated in both the cases mentioned above can be corrected using a strategy similar to the one adopted in the construction of a fault-tolerant logical \textsc{cz} gate.

\section{Universal set of logical gadgets}\label{sec:universal}

We are now ready to construct a universal set of logical gadgets, using the basic fault-tolerant components of the previous section, tailored for amplitude-damping noise. In particular, our universal set of gadgets comprise
\begin{enumerate}
\item preparation of the $|0\rangle_L$ and $|+\rangle_L$ states;
\item measurement of logical $X$ and $Z$;
\item two-logical-qubit gate \textsc{cz};
\item single-logical-qubit gate $H$, $S$, and $T$.
\end{enumerate}
Here, the logical $H$ is the Hadamard gate, $\overline H\equiv |+\rangle_L\langle 0|+|-\rangle_L\langle 1|$; the logical $S$ is the phase gate, $\overline S\equiv |0\rangle_L\langle 0|+\mi |1\rangle_L\langle 1|$; and the logical $T$ (or $\pi/8$) is the gate $\overline T\equiv |0\rangle_L\langle 0| + \mathrm{e}^{\mi\pi/4}|1\rangle_L\langle 1|$. These gadgets can be strung together to form fault-tolerant computational circuits. The fault-tolerant constructions of the $\overline X$ and $\overline Z$ measurement gadgets are described in SM Sec.~B, whereas the $\textsc{cz}$ gadget is already given in Sec.~\ref{sec:enc_gadget}. The preparation gadgets for $|0\rangle_L$ and $|+\rangle_L$ are also straightforward to construct using a Bell-state preparation gadget, as described in SM Sec.~C. It remains then to describe the fault-tolerant construction of the logical $H$, $S$, and $T$ gadgets.

As was the problem with the \textsc{cnot} gate, the physical $H$, $S$ and $T$ gates are not noise-structure preserving: They change an input damping error into an error not correctable by the $4$-qubit code. We thus do not have transvsersal implementations of these logical gates; rather, we need a different approach for getting fault-tolerant logical gate operations. Here, we make use of the well-known technique of gate teleportation~\cite{knill_FT,nielsen} to construct our fault-tolerant logical gadgets. The resulting logical gadgets are manifestly fault-tolerant against amplitude-damping noise as we build the teleportation circuits using the basic encoded gadgets shown to be fault-tolerant in Sec.~\ref{sec:enc_gadget}.

\begin{figure}[H]
\begin{center}
\includegraphics[width=0.6\columnwidth]{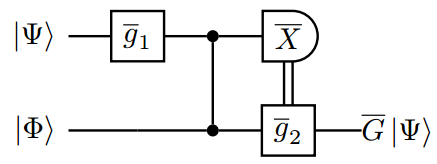}
\caption{Single-logical-qubit gadget implementing the gate $\overline G=\overline H, \overline S,$ or $\overline T$. $|\Psi\rangle$ is the incoming state on which $\overline G$ is to be applied. For $\overline G=\overline H$, $|\Phi\rangle=|+\rangle_L$, $\overline g_1$ is the identity, and $\overline g_2=\overline X$; for $\overline G=\overline S$, $|\Phi\rangle=|\Phi_S\rangle$, $\overline g_1=\overline H$, and $\overline g_2=\overline Y=\overline{XZ}$; for $\overline G=\overline T$, $|\Phi\rangle=|\Phi_T\rangle$, $\overline g_1=\overline H$, and $\overline g_2=\overline{SX}$. The gate $\overline g_2$ is applied conditioned on the outcome of the $\overline X$ measurement.
}
\label{fig:HST}
\end{center}
\end{figure}

Specifically, all three single-logical-qubit gadgets share the same teleportation structure shown in Fig.~\ref{fig:HST}. Each require a resource state---$|+\rangle_L$ for $\overline H$, $|\Phi_S\rangle\equiv\frac{1}{\sqrt{2}}{\left(\ket{0}_L + \mi\ket{1}_L\right)}$ for $\overline S$, and $|\Phi_T\rangle \equiv \frac{1}{\sqrt{2}}{\left(\ket{0}_L + \mathrm{e}^{\mi\pi/4}\ket{1}_L\right)}$ for $\overline T$---as input, the use of the logical $X$ measurement gadget, as well as an additional logical gadget, conditioned on the measurement outcome, to complete the teleportation. $\overline S$ and $\overline T$ also require the use of the $\overline H$ gadget. The definition and fault-tolerant preparation of the resource states $|\Phi_S\rangle$ and $|\Phi_T\rangle$ can be found in SM Sec.~C. That these gadgets are fault tolerant follows simply from the fact that their components are the fault-tolerant gadgets of Sec.~\ref{sec:enc_gadget}.


\section{Pseudothreshold calculation}\label{sec:threshold}

The feasibility of any fault tolerance scheme is partly quantified by its error threshold, which refers to that critical value of the noise strength, $p_{\text{th}}$, at which the fault tolerance protocol fails to outperform the unencoded version. In this work, we use the infidelity metric $\mathrm{IF}( . , .)$, 
\begin{equation}
\mathrm{IF}(\rho,\sigma)\equiv1-{\left(\tr\sqrt{\rho^{1/2}\sigma\rho^{1/2}}\right)}^2,
\end{equation}
for a pair of states $\rho$ and $\sigma$, to benchmark the performance of the protocol. For a physical (i.e., unencoded) noise strength $p$, if the fault tolerance procedure can correct up to one error, then (see SM Sec.~D1) the infidelity of the output state of a fault-tolerant gadget with respect to the ideal output is upper-bounded by $Cp^2 + Bp^3$, assuming a pure input state. Here, $C$ is the number of \emph{malignant fault pairs}---pairs of faults in the fault-tolerant gadget that propagate into an uncorrectable output---while $B$ refers to the number of ways in which the gadget can have a third-order fault.

The critical noise threshold $p_\text{th}$ for a given initial state is then lower-bounded by the $p\equiv p_\text{th}^{(l)}$ that solves
\begin{equation}\label{eq:lower_threshold}
	Cp^2 + Bp^3 = \mathrm{IF}_{p}(\rho, \tilde{\rho}),
\end{equation}
where $\mathrm{IF}_{p}(\rho, \tilde{\rho})$ is the infidelity, for the unencoded version of the gadget, between its noisy output state $\tilde{\rho}$ and the ideal output $\rho$. The subscript $p$ emphasizes the dependence of the infidelity on the physical noise strength $p$.
Eq.\eqref{eq:lower_threshold} compares the encoded scheme to the unencoded one, and we refer to the resulting threshold $p_\text{th}$ as a \emph{pseudothreshold}, following Refs.~\cite{cross2009,napp2012optimal}. This is different from the usual quantum accuracy threshold discussed in concatenated-code fault tolerance treatments which requires a recursive simulation argument to go to higher levels of encoding for increased error-removing power (see, for example, Ref.~\cite{aliferis}).

We note that, in general, the pseudothreshold $p_\text{th}$ as well as the bound $p_\text{th}^{(l)}$ depend on the input state as the unencoded infidelity $\mathrm{IF}_{p}(\rho, \tilde{\rho})$ varies with the input state. One reasonable way to obtain a state-independent measure is then to report the mean values of the pseudothreshold and pseudothreshold bounds over all pure states, denoted as $\overline{p}_\text{th}$ and $\overline{p}_\text{th}^{(l)}$, respectively. 

We also note that for amplitude damping noise, apart from malignant fault pairs, one must include in $C$, order-$p^2$ $Z$ errors arising from a single fault [see Eq.~\eqref{eq:ampdamp2}]. Moreover, one must be careful in assigning weights to each fault pair in order to obtain a tight bound. Recall that a fault $\cF$ can cause two kinds of errors, $\cF_a$ and $\cF_z$, and not all the combinations of $\cF_a$ and $\cF_z$ lead to an uncorrectable output. For example, two \textsc{cnot}s in a parity measurement [see Fig.~\ref{fig:basic_components}(a)] each can have one fault at the controls, one with $\cF_a$ error and the other with $\cF_z$ error, and the output is still correctable; however, if the two faults cause two $\cF_a$ errors, then the output has a logical error. Moreover, in case the two faults are both $\cF_z$, the multiplicative factor should be $p^2/4$ [see Eq.~\eqref{eq:amp_fault}] instead of $p^2$. It is also often the case that a pair of locations with two $\cF_a$ errors can lead to an output with an $X$ error or an $(I\pm Z)$ error on one of the four data qubits. In such cases, we still get a correct state half of the time when trying to correct the output, and therefore, the multiplicative factor should be $p^2/2$. Taking all of these factors into account, we obtain a better estimate of $C$, leading to tighter bounds for the pseudothreshold.

An ideal circuit is simulated fault tolerantly by replacing each unencoded gadget in the circuit with an encoded gadget followed by an \textsc{ec} gadget. The failure probability of such a fault-tolerant simulation can be expressed in terms of the failure probability of overlapping composite objects constituting the circuit, called \emph{extended gadgets}, which take into account both incoming errors and faults occurring within a given gadget. Therefore, an extended gadget often includes both the leading \textsc{ec} gadget and the trailing \textsc{ec} gadget with the encoded gadget sandwiched in between. The fault tolerance of the simulation circuit is then ensured by the fault tolerance of the extended gadgets. We refer the readers to Ref.~\cite{aliferis} for a detailed discussion of this argument. In this section, we obtain the pseudothresholds for two situations, namely, the memory circuit with no non-trivial computational operations, and a general computational circuit.

\subsection{Memory pseudothreshold}\label{sec:mem_threshold}

 \begin{figure}[H]
\centering
\includegraphics[scale=.4]{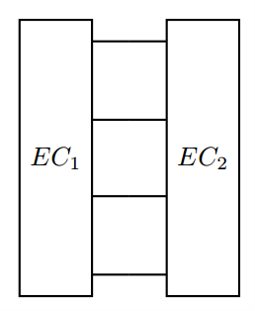}
\caption{Extended identity gadget.}
\label{fig:memory_unit}
\end{figure} 
Suppose we are interested only in storing the quantum information for a certain period of time. This can be thought of as a circuit comprising only identity (i.e., trivial) computational gates, with periodic error correction operations to remove errors and thus preserve the stored information. The relevant extended gadget comprises a pair of \textsc{ec} gadgets surrounding the identity gate, corresponding to storage for the time between consecutive error correction cycles, as shown in the Fig.~\ref{fig:memory_unit}. To obtain the pseudothreshold for this memory situation, we enumerate the number of malignant pairs of faults leading to an uncorrectable output, assuming that the incoming state has no errors. We label the different blocks that constitute the identity gadget as follows:
\begin{enumerate}
	\item \textsc{ec$_1$} - leading \textsc{ec}
	\item \textsc{ec$_2$} - trailing \textsc{ec}
	\item $4$ rest locations
\end{enumerate}
We can then represent the number of malignant pairs via a matrix whose rows and columns correspond to each block in Fig.~\ref{fig:memory_unit}, with the entries of the matrix denoting the total malignant-pair contributions from the respective blocks \cite{aliferis}. Because of the overlap between two consecutive extended gadgets when they are strung up into the memory circuit, to avoid double counting, a fault pair in the leading \textsc{ec} of an extended gadget is counted as a fault pair in the trailing \textsc{ec} of the preceding extended gadget. 
\[
\begin{blockarray}{ccccc}
	&  & 1 & 2 & 3 \\
	\begin{block}{cc(ccc)}
		1 & & 0 & & \\
		2 & & 702 & 5542 & \\
		3 & & 28 & 224 & 6 & \\
	\end{block}	
\end{blockarray}
\]
The total number of malignant pairs -- $6502$ -- is simply the sum of the entries of the matrix. Apart from the pairs of damping faults, we also need to keep track of the $O(p^{2})$ phase errors, and we argue in SM Sec.~D2 that there are $29$ malignant fault locations for the phase errors. In total, we obtain $C=6531$ and $B= 8,171,621$, leading to the average bound,
\begin{equation}\label{eq:mem_thres}
\overline{p}_{\rm th}^{(l)} = 5.13\times10^{-5}.
\end{equation}
We refer the reader to SM Sec.~D2 for the details of the calculation.

\subsection{Computational pseudothreshold}\label{sec:comp_threshold}

\begin{figure}[H]
\centering
\includegraphics[scale=0.4]{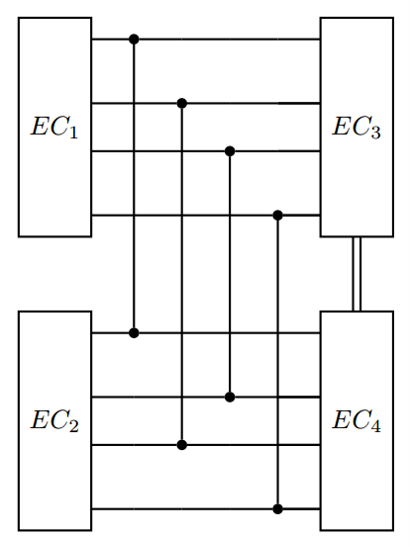}
\caption{Extended \textsc{cz} gadget for computing the pseudothreshold}
\label{fig:exrec}
\end{figure}

Next, we consider a general circuit, comprising a sequence of computational gates, chosen from the universal logical gate set of Sec.~\ref{sec:universal}. Among all the possible extended gadgets constructed from our set of basic encoded gadgets, the extended \textsc{cz} gadget, shown in Fig.~\ref{fig:exrec}, turns out to have the maximum number of malignant pairs, as verified by exhaustive counting. This \textsc{cz} gadget thus determines the pesudothreshold relevant for this computational situation. 

Similar to the extended identity gadget, we label the different blocks that constitute the extended \textsc{cz} as follows:
\begin{enumerate}
	\item \textsc{ec$_1$} - leading \textsc{ec} in block 1
	\item \textsc{ec$_2$} - leading \textsc{ec} in block 2
	\item \textsc{ec$_3$} - trailing \textsc{ec} in block 1
	\item \textsc{ec$_4$} - trailing \textsc{ec} in block 2
	\item \textsc{cz} gadget
\end{enumerate}
We can then represent the number of malignant pairs via the following matrix. We merely note the final numbers here and refer to SM Sec.~D3 for the detailed enumeration of malignant pairs for every pair of blocks: 
\[
\begin{blockarray}{ccccccc}
	&  & 1 & 2 & 3 & 4 & 5 \\
	\begin{block}{cc(ccccc)}
		1 & & 0 & & & & \\
		2 & & 48 & 0 & & & \\
		3 & & 718 & 328 & 5542 & &  \\
		4 & & 328 & 718 & 24 & 5542 & \\
		5 & & 28 & 28 & 230 & 230 & 12 \\
	\end{block}	
\end{blockarray}
\]
We obtain $C = 13835$ and $B= 65,371,138$, leading to the average computational pseudothreshold lower-bounded by
\begin{equation}
\overline{p}_{\rm th}^{(l)} = 2.26\times 10^{-5}.
\end{equation}

\subsection{Simulating the pseudothreshold}\label{sec:sim}

Finally, as a check on our counting analysis of the previous subsections, we present the memory pseudothreshold obtained from a computer simulation of the extended memory gadget of Fig.~\ref{fig:memory_unit}. For a given initial state $\rho_\mathrm{in}$ and a physical noise strength $p$, the \textsc{ec} circuit (Fig.~\ref{fig:ec}) is simulated, with the amplitude-damping channel applied to every qubit at every time step. The intermediate measurements and recovery operations are applied as quantum channels so we obtain a single density matrix at the end of a simulation. This state is passed through an ideal decoder (see Sec.~\ref{sec:prelim} and SM Sec.~D4) to strip off $O(p)$ errors and project the state to the code space, before comparing it with the ideal output state $\rho_0$, which, for a memory gadget, coincides with the initial state $\rho_\mathrm{in}$. The pseudothreshold is determined as the physical noise strength at which the encoded infidelity between $\rho_0$ and the final output state starts becoming smaller than the unencoded infidelity between the same initial state encoded in a single qubit and the output after passing it through an amplitude-damping channel.

\begin{figure}
\centering
\includegraphics[clip, width=\columnwidth]{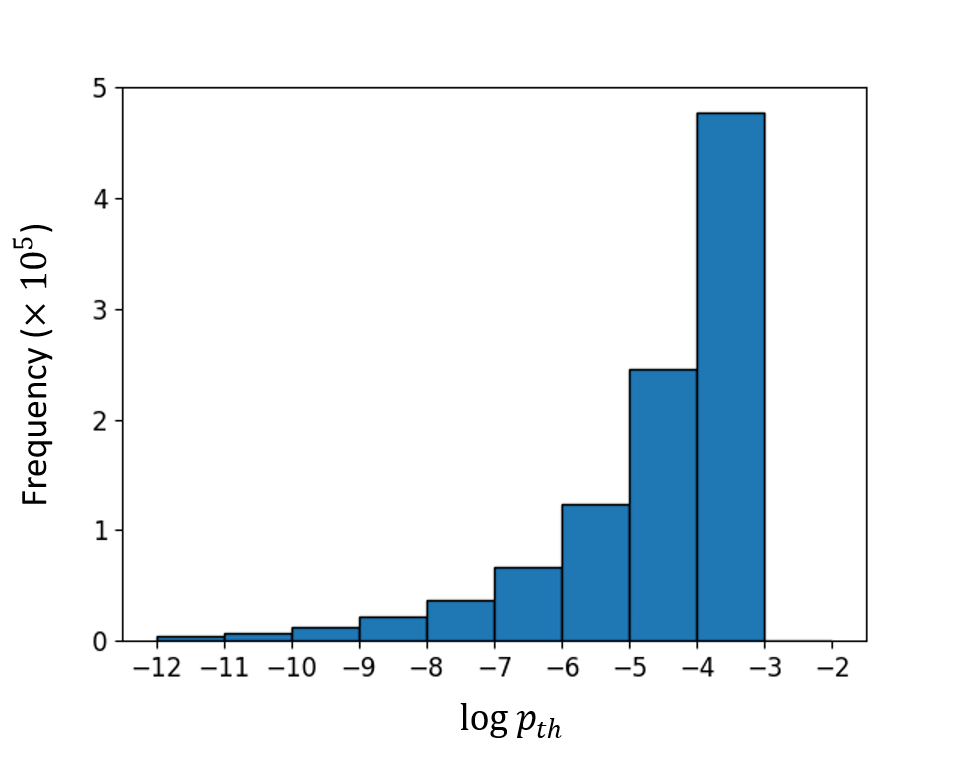}
\caption{Histogram of simulated pseudothresholds. The bins are in intervals of the logarithm of $p_{\rm{th}}$, and the vertical axis plots frequencies in hundred thousands ($10^5$).}
\label{fig:memory_sim}
\end{figure}

As noted in Sec.~\ref{sec:threshold}, this pseudothreshold is very state dependent, as the infidelity measure itself is state dependent. For example, states close to the fixed state of the amplitude-damping noise ($|0\rangle$) achieve high fidelity even in the unencoded case. In Fig.~\ref{fig:memory_sim}, we plot the distribution of pseudothresholds for $N=10^6$ randomly sampled pure initial states. We indeed see that we get a distribution of the simulated pseudothresholds, and one that is in fact heavily skewed. The mean pseudothreshold value is $1.56\times 10^{-4}$, which is higher than the bound $\overline{p}_{\rm th}^{(l)} =2.98\times 10^{-5}$ obtained from the counting method of Sec.~\ref{sec:mem_threshold}, verifying that our counting gives a valid lower bound on the infidelity. Note that this $\overline{p}_\mathrm{th}^{(\ell)}$ value differs slightly from that in Eq.~\eqref{eq:mem_thres}, to account for differences between the simulation and counting procedures; see SM Sec.~D4 for further details).

\section{Conclusion}\label{sec:concl}

We demonstrate a universal fault tolerance scheme tailored to amplitude-damping noise, using the $4$-qubit code. We construct an error correction gadget, preparation gadgets, measurement gadgets, and an encoded universal gate set, all tolerant to single-qubit faults arising due to amplitude-damping noise. Our construction shows that achieving fault tolerance using noise-adapted codes for non-Pauli noise models like amplitude-damping is possible, but poses interesting challenges, and can lead to counter-intuitive results when viewed from the standpoint of the well-established principles of quantum fault tolerance. 

For instance, our \textsc{ec} gadget requires a nontrivial recovery unit to correct for the single-qubit off-diagonal fault $\cF_{z}$, unlike the standard fault tolerance schemes that only require a simple classical Pauli frame change. Another significant departure from the conventional ideas of fault tolerance is the fact that logical gates such as the \textsc{cnot} and the logical $\overline{X}$ which are transversal, are however not fault-tolerant against single-qubit damping errors. This in turn motivates the need to identify \emph{noise-structure preserving} gates while developing fault-tolerant schemes using noise-adapted codes.
Indeed, the structure of the non-Pauli noise dictates our choice of fault-tolerant gate gadgets. Thus, the transversal two-qubit \textsc{cz} gate turns out to be a more natural choice for a two-qubit gate, rather than the transversal \textsc{cnot} gate. When it comes to single-qubit gates, we do not obtain any transversal constructions for the $4$-qubit code. Rather, we have to rely on gate teleportation to implement the Hadamard, $S$ and $T$ gates. These additional complications contribute to the perhaps poorer-than-expected pseudothreshold for the encoded gadgets.

Our work presents a first step towards achieving fault tolerance against specific noise models, and can already be used as an initial noise-reduction step towards more accurate computation. A further step would be to investigate possibilities of optimizing the gadget constructions for smaller ones with fewer fault locations and hence better pseudothreshold. One could even ask the standard fault tolerance question of scaling up the code beyond a single layer of encoding, by concatenation for example, or via the $2D$ Bacon-Shor code~\cite{piedrafita2017reliable} generalization of the 4-qubit code. Such extensions of our work could provide more error correction power even within a resource-constrained scenario and have the potential to take us closer to more accurate---and hence more useful---quantum computers.

\begin{acknowledgments}
	This work is supported in part by the Ministry of Education, Singapore (through grant number MOE2018-T2-2-142). HKN also acknowledges support by a Centre for Quantum Technologies (CQT) Fellowship. CQT is a Research Centre of Excellence funded by the Ministry of Education and the National Research Foundation of Singapore. PM acknowledges financial support by the Department of Science and Technology, Govt. of India, under grant number DST/ICPS/QuST/Theme-3/2019/Q59. This work is supported in part by a seed grant from the Indian Institute of Technology Madras, as part of the Centre for Quantum Information, Communication and Computing. 

\end{acknowledgments}

\bibliography{FTAmpDamp}

\clearpage

\begin{center}
	{\large \textbf{Supplemental Material}}
\end{center}

\setcounter{section}{0}
\renewcommand{\thesection}{\Alph{section}}
\renewcommand{\thesubsection}{\arabic{subsection}}

\section{Principles of fault tolerance}\label{app:faulttolerance}

We formally state the properties of fault-tolerant gadgets here. Specifically, we list the properties that the error correction gadget and the encoded gadgets must satisfy, in order to lead to logical operations and circuits that are fault-tolerant against amplitude-damping noise. These properties are used in the proofs of fault tolerance in SM Sec.~\ref{app:basic_units}. 
\begin{itemize}
	\item[(P1)] If an error correction gadget has no fault, it takes an input with at most one  error to an output with no errors.
	\item[(P2)] If an error correction gadget contains at most one fault, it takes an input with no errors to an output with at most one error.
	\item[(P3)] A preparation gadget without any fault propagates an input with up to one error to an output with at most one error. A preparation gadget with at most one fault propagates an input with no errors to an output with at most one error. 
	\item[(P4)] A measurement gadget with no faults leads to a correctable classical outcome for an input with at most one error. A measurement gadget with at most one fault anywhere leads to a correctable classical outcome for an input with no errors. 
	\item[(P5)] An encoded gadget without any fault takes an input with up to a single error to an output in each output block with at most one error. An encoded gadget with at most one fault takes an input with no error to an output with at most one error in each output block.
\end{itemize}

\section{Fault tolerance analysis of basic gadgets}\label{app:basic_units}
In this appendix, we give the full constructions of the Bell-state preparation gadget and the $\overline X$ and $\overline Z$ measurement gadgets mentioned in Sec.~\rom{4} of the main text. In addition, we provide the details of the fault tolerance analysis of the \textsc{ec} gadget, as well as of the logical $X$, $Z$, and $\textsc{cz}$ operations described in Sec.~\rom{4}.

\subsection{Bell-state preparation}\label{sec:prep}

\begin{figure}
	\includegraphics[trim=35mm 170mm 30mm 0mm, clip, width=0.35\columnwidth]{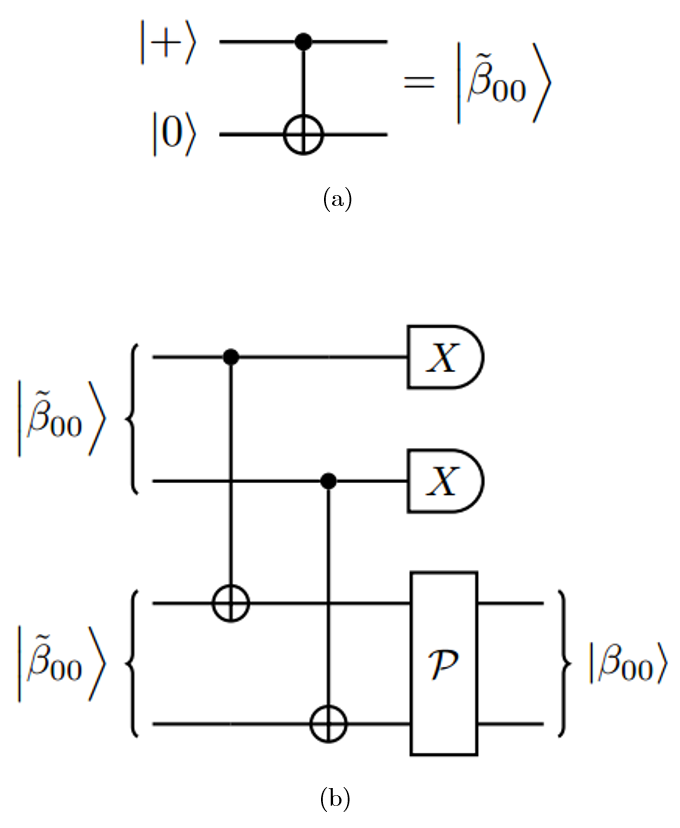}
	\hspace*{0,5cm}
	\includegraphics[trim=5mm 15mm 0mm 80mm, clip, width=0.5\columnwidth]{Figure8-BellPrep.png}\\
	\hspace*{-1cm}(a)\hspace*{4cm}(b)
	\caption{\label{fig:BellPrep}Preparation of the Bell state: (a) standard, non-fault-tolerant circuit for the preparation of the two-qubit Bell state $|\tilde{\beta}_{00}\rangle$; (b) a verification circuit for a fault-tolerant preparation of $|\beta_{00}\rangle$, using two copies of $|\tilde{\beta}_{00}\rangle$ created from the circuit in (a).}
\end{figure}

We describe a fault-tolerant preparation of the two-qubit Bell state,
\begin{equation}\label{eq:bell}
	|\beta_{00}\rangle = \frac{1}{\sqrt 2}(|00\rangle+|11\rangle),
\end{equation}
which serves as the input state to multiple fault-tolerant gadgets constructed in SM Sec.~\ref{sec:logical_prep}.
The Bell state is first prepared in a non-fault-tolerant manner, denoted as $|\tilde{\beta}_{00}\rangle$ in Fig.~\ref{fig:BellPrep}(a). We then verify this Bell state using another copy of $|\tilde{\beta}_{00}\rangle$ prepared in a similar fashion, as shown in Fig.~\ref{fig:BellPrep}(b). 
The Bell state is accepted for use in further computation if the $X$ measurements yield even parity, i.e., both are $0$ or both are $1$, and, the parity measurement gives a $0$ outcome; otherwise, the final state is rejected and we start over.

That Figs.~\ref{fig:BellPrep}(a) and (b) lead to a fault-tolerant preparation of $|\beta_{00}\rangle$---satisfy property (P3) of SM Sec.~\ref{app:faulttolerance}---can be understood as follows.
First, we consider an $\cF_z$ error anywhere in the circuits. An $\cF_z$ occurring before or at the control of two $\textsc{cnot}$s used for the $X$-measurements is killed of by the $X$-measurements themselves. An $\cF_z$ at the target of those two $\textsc{cnot}$s or in the parity measurement leads to at most one $\cF_z$ error in the accepted Bell state.

Next, we consider the effect of a damping error $\cF_a$.
\begin{itemize}
	\item A faulty Hadamard results in the preparation of the state $\ket{00}$ by the circuit in Fig.~\ref{fig:BellPrep}(a). If this happens to the first block of $\ket{\Tilde{\beta}_{00}}$ in the circuit in Fig.~\ref{fig:BellPrep}(b), it has no effect on the second block but may change the outcomes of two $X$-measurements. The second Bell state is still rejected if the outcomes have odd parity. On the other hand, if the faulty Hadamard is in the second Bell state block, only even parity outcomes correspond to a correct Bell state; odd parity outcomes correspond to the state $\ket{00} - \ket{11}$ in the second block, which is rejected.
	\item A fault in the \textsc{cnot} in Fig.~\ref{fig:BellPrep}(a), at either the control or target, leads to an odd outcome for the parity measurement at the end. A fault at the control of the \textsc{cnot}s in Fig.\ref{fig:BellPrep}(b) or a faulty $X$-measurement has the same effect as a faulty Hadamard in the first block: it may cause odd parity outcomes of two $X$-measurements but does not affect the second block. On the other hand, an fault at the target causes an odd outcome for the parity measurement. 
	
	\item Finally, a fault in the parity measurement is either detected by the parity measurement itself, or causes at most one error to the outgoing Bell state, in case it is accepted.
\end{itemize}

\subsection{Logical $Z$ and $X$ measurements}\label{sec:X_meas}

Next, we demonstrate fault-tolerant circuits that perform logical $\overline X$ and $\overline Z$ measurements corresponding to the $4$-qubit code. Recall [property (P4)] that a measurement gadget is said to be fault-tolerant if the presence of a single fault in the measurement circuit always leads to a correctable error in the classical outcome. In other words, distinct faults lead to distinct classical outcomes, so as to ensure that the faults can be diagnosed and corrected for, classically.

We first argue that the logical $Z$, i.e., $\overline{Z} = ZIZI$, measurement can be realised simply by performing four local $Z$ measurements on the encoded (data) qubits. In the ideal, no-error scenario, the measurement outcomes (which are four classical bits) have even parity. Specifically, the outcomes $0000$ and $1111$ correspond to the data qubits projected onto the $|0\rangle_{L}$ state, whereas the strings $0011$ and $1100$ correspond to the data qubits projected onto $|1\rangle_{L}$. A single $\cF_z$ error has no effect since it commutes with a $Z$-measurement. However, a single damping error $\cF_a$ in one of the local $Z$ measurements, or in the data qubits, leads to outcomes with odd parity. Furthermore, faults in distinct locations lead to distinct four-bit classical strings, thus ensuring that the faults can be diagnosed and corrected for. In particular, if the outcome string has more $1$'s, then the \emph{correct} outcome corresponds to a projection onto the $|0\rangle_{L}$ state, whereas if the outcome string has more $0$'s the correct outcome corresponds to a projection onto the $|1\rangle_{L}$ state.

\begin{figure}
	\includegraphics[width=0.45\columnwidth]{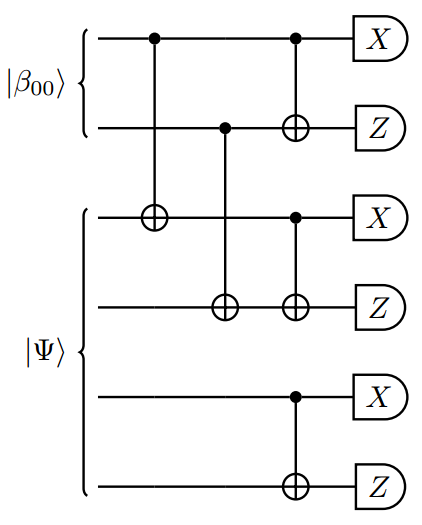}
	\caption{\label{fig:x_meas} Measurement of the logical operator $\overline{X}$=$XXII$ on the data qubits (the bottom $4$ qubits) with the use of two additional ancilla qubits (top $2$ qubits) fault-tolerantly prepared in $|\beta_{00}\rangle$.}
\end{figure}

Next, we give a fault-tolerant construction of the logical $X$ --- $\overline{X}=XXII$ --- measurement gadget, by introducing two additional ancilla qubits prepared in the Bell state $|\beta_{00}\rangle$, as shown in Fig.~\ref{fig:x_meas}. The first two qubits are ancilla qubits and the bottom four are data qubits. We perform three Bell measurements, one on the two ancilla qubits, and two each on each pair of data qubits. The Bell measurements on an \emph{ideal} encoded state (an encoded state with no error) lead to outcomes $(0,0)$ or $(1,0)$. Outcome $(0,0)$ indicates a projection onto $|+\rangle_{L}$, and the outcome $(1,0)$ indicates a projection onto $|-\rangle_{L}$, thus realising the logical $X$ measurement. 

This $\overline X$ measurement gadget can be shown to be fault tolerant, that is, it leads to a correctable classical outcome even when there is a single fault anywhere in the circuit. We do that next, but as an intuitive reason for fault tolerance, we note that a single fault in the circuit would lead to a \emph{faulty} outcome ($(1,1)$ or $(0,1)$) for one of the Bell measurements. We simply ignore the faulty outcome and choose the majority among the rest of outcome pairs ($(0,0)$ and $(1,0)$) to decide if the state of the data qubits is projected onto $|+\rangle_{L}$ or $|-\rangle_{L}$.

We now prove that the $\overline{X}$ measurement gadget satisfies the fault tolerance property (P5) stated in SM Sec.~\ref{app:faulttolerance}.
Assuming the initial state of the data qubits is $\ket{\Psi} = a\ket{+}_L + b\ket{-}_L$, then after the two \textsc{cnot}s between the ancilla qubits at the top and the four (data) qubits in Fig.~\ref{fig:x_meas}, the state is given by,

\begin{equation*}
	a \ket{\beta_{00}} ^{\otimes 3} + b \ket{\beta_{10}} ^{\otimes 3}
\end{equation*}
where the first pair refers to the ancilla qubits and the other two pairs refer to the data qubits. Apart from $\ket{\beta_{00}}$, other Bell states are also labeled according to their outcome in a Bell measurement, namely, $\ket{\beta_{10}} = \tfrac{1}{\sqrt{2}}(\ket{00}-\ket{11})$, $\ket{\beta_{01}} = \tfrac{1}{\sqrt{2}}(\ket{01}+\ket{10})$, $\ket{\beta_{11}} = \tfrac{1}{\sqrt{2}}(\ket{01}-\ket{10})$.

If there is no error in the incoming state as well as no fault in the measurement circuit, then the outcomes of three Bell measurements are either all $(0,0)$ or $(1,0)$, corresponding to the data qubits projected to $\ket{+}_L$ or $\ket{-}_L$. However, if there is error or fault the outcomes may be different. First, note that an $\cF_z$ error in any data qubits or ancilla qubits, or an $\cF_z$ fault in any location of the measurement circuit is killed off by at least one of three Bell measurements. This can be seen by commuting a $Z$ error through the measurement circuit and noticing that it always meets at least one $X$-measurement. Therefore, we only need to consider the damping error $\cF_a$.

Now, we consider an error at one of the data qubits. 
\begin{itemize}
	\item If data qubit 1 or 2 is damped, then the state after being entangled with the ancilla pair is 
	\begin{equation*}
		(\ket{\beta_{00}}\ket{\beta_{01}} \pm \ket{\beta_{10}}\ket{\beta_{11}})(a\ket{\beta_{00}} -b\ket{\beta_{10}})
	\end{equation*}
	If the outcomes of the Bell measurements are $\{(00), (01), (00)\}$ or $\{(10), (11), (10)\}$, we can discard the invalid outcome $(01)$ and $(11)$ and determine the correct outcome of the $\Bar{X}$ measurement based on the other two. However, if the outcomes are $\{(00), (01), (10)\}$ or $\{(10), (11), (00)\}$, then there is a tie after discarding the invalid outcome. In this case, we further discard the outcome of the ancilla block and determine the correct outcome based on the third Bell measurement. The reason for this is that the outcomes of the first and second Bell measurements can be both wrong due to the coupling of two $\textsc{cnot}$s, whereas the third Bell measurement is independent from the other two.
	\item If data qubit 3 or 4 is damped, then the state after entangled with the ancilla block is
	\begin{equation*}
		(a\ket{\beta_{00}}\ket{\beta_{00}} -b\ket{\beta_{10}}\ket{\beta_{10}})(\ket{\beta_{01}} \pm \ket{\beta_{11}})
	\end{equation*} 
	In this case, the outcome of the third Bell measurement is discarded; the correct outcome is determined based on the other two Bell measurements. 
\end{itemize}

Next, we consider a fault in the measurement circuit. 
\begin{itemize}
	\item If one of two ancilla qubit is faulty, then the state after entangled with the data qubits is
	\begin{equation*}
		(\ket{\beta_{01}} \pm \ket{\beta_{11}})(a\ket{\beta_{01}}\ket{\beta_{00}} \pm b\ket{\beta_{11}}\ket{\beta_{10}})
	\end{equation*}
	For this case, after discarding invalid outcomes, the last one gives the correct result. 
	\item If one of two \textsc{cnot}s used to entangle the data qubits with the ancilla qubits is faulty, either at control or target, it is easily to verify that the state right before the Bell measurements is one of the following states
	\begin{equation*}
		\begin{split}
			& (\ket{\beta_{01}} \pm \ket{\beta_{11}})(a\ket{\beta_{00}}\ket{\beta_{00}} - b\ket{\beta_{10}}\ket{\beta_{10}}) \\
			& a\ket{\beta_{00}}(\ket{\beta_{01}} \pm \ket{\beta_{11}})\ket{\beta_{00}} - b\ket{\beta_{10}}(\ket{\beta_{01}} \pm \ket{\beta_{11}})\ket{\beta_{10}}
		\end{split}
	\end{equation*}
	As the above case, the correct logical outcome is determined after discarding the invalid Bell measurement outcomes. 
	\item Finally, a fault in one of three Bell measurements may spoil the outcome of that measurement. However, the other two are unaffected and we can correctly determine the outcome of $\overline{X}$ measurement from those two.
\end{itemize}

This covers all the possibilities of a single fault in the measurement circuit in Fig.~\ref{fig:x_meas}, hence, shows that the $\overline{X}$ measurement is fault tolerant.

\subsection{\textsc{ec} gadget}\label{app:ECFT}

We want to show that the \textsc{ec} gadget described in Sec.~\rom{4}A of the main text is fault tolerant in that it has properties (P1) and (P2) of SM Sec.~\ref{app:faulttolerance}.
That (P1) holds is ensured by the fact that the syndrome extraction and recovery, when without fault, can correct up to one damping error. We notice that an incoming state without error of the form $\ket{\Psi} = a\ket{0}_L + b\ket{1}_L$ becomes
\begin{equation}
	\label{eq:entangled_state}
	\begin{split}
		\ket{\tilde{\Psi}} = &\tfrac{a}{\sqrt{2}}(\ket{0000}\ket{0000}_f + \ket{1111}\ket{1111}_f) \\
		+&\tfrac{b}{\sqrt{2}}(\ket{0011}\ket{0011}_f + \ket{1100}\ket{1100}_f) ,
	\end{split}	
\end{equation}  
after entangling with the flag qubits by the first four \textsc{cnot}s (see Fig.~1 of the main text), where the subscript $f$ denotes the set of four flag qubits. When there are no faults, the state right before the four $Z$-measurements is $\ket{\Psi}\ket{0000}_f$. As a result, measurement of the flag qubits will give the outcome $0000$ and leave the data qubits in state $\ket{\Psi}$. On the other hand, if there is a damping error in one of the data qubits, either the syndrome extraction unit will detect the damping part $\cF_a$ and the recovery $\cR$ will correct for it, or the syndrome extraction unit will detect no damping, let the off-diagonal part $\cF_z$ through, which is then killed off by the $\cM_1$ unit. The fact that the $\cM_1$ unit can kill off the $\cF_z$ error can be seen by noting that the state in Eq.~\eqref{eq:entangled_state} is stabilized by $XXXXXXXX$. Therefore, the projection on the $+1$ eigenspace of that state with an off-diagonal $Z$ is 
\begin{equation*}
	\begin{split}
		&\tfrac{1}{2}(1+X^{\otimes 8}) (Z_i\ket{\tilde{\Psi}} \bra{\tilde{\Psi}}) \tfrac{1}{2}(1+X^{\otimes 8}) \\
		&= Z_i\tfrac{1}{2}(1-X^{\otimes 8}) \ket{\tilde{\Psi}} \bra{\tilde{\Psi}} \tfrac{1}{2}(1+X^{\otimes 8}) = 0 .
	\end{split}
\end{equation*}
In any case, the outgoing state has no error as promised.

To verify (P2), we need to consider faults at different locations in the \textsc{ec} gadget shown in Fig.~1 of the main text.
We recall that a faulty \textsc{cnot} is modeled as an ideal \textsc{cnot} followed by a fault $\cF$ on both the control and target qubits; a faulty measurement is modeled as a fault $\cF$ followed by an ideal measurement. Since the off-diagonal part $\cF_z$ passes through all the parity measurements and is only killed off by the $\cM_1$ unit, a $\cF_z$ error occurring in any location before the $\cM_1$ unit will not survive, whereas a $\cF_z$ occurring inside or after the $\cM_1$ unit will lead to one $\cF_z$ error at the output. Hence, (P2) is satisfied for the $\cF_z$ error and from now on, we only consider the effect of $\cF_a$ error.

A fault at the control of one of the first four \textsc{cnot}s is detected by the following syndrome extraction and is corrected by the recovery unit $\cR$. On the other hand, a fault at the target, denoted as $D_i$ ($i=1,2,3,4$) (see Tab.~\ref{tab:R0syn}), causes a flip in the corresponding flag qubit and possibly propagates a $Z$ error to the data qubits, depending on the outcome of $\cM_1$. Note that the flag syndrome alone is not enough to conclude that the fault is at the flag qubits because a fault in a data qubit may cause the same syndrome, as discussed later. Hence, another parity measurement is performed and we can conclude that the fault is at the flag qubit if the parity measurement gives trivial outcome. Now, the $Z$ error is taken care of by a measurement of $XXXX$ on the data qubits and a $Z$ is applied correspondingly if the outcome is $1$.

Next, let us consider the syndrome extraction unit $\cS$. Note that unless the syndrome bits $s_1$ or $s_2$ are triggered, i.e., record a 1, neither of the subsequent gates in the syndrome extraction unit that measures $ZIII$, $IZII$, $IIZI$, or $IIIZ$, will be performed. $s_1$ and $s_2$ are not triggered, assuming no incoming errors, unless a fault occurs in the gates that perform those parity measurements, namely, at two $\textsc{CNOT}$ locations in Fig.~2(a) of the main text. Note that faults can also occur in the ``resting" locations within the same time-steps, but those can all be grouped into either incoming errors for this syndrome extraction unit, or undetected errors that will be fixed only by the next part of the \textsc{ec} gadget. We list below the possible faults at different locations in the syndrome unit, and explain how they are diagnosed.

\begin{itemize}
	\item A faulty \textsc{cnot} at location $1$ in Fig.~2(a) of the main text involves two cases: (i) fault at control; (ii) fault at target. Case (i) is not detected in this parity measurement, and will present as an outgoing damping error, to be dealt with in the next part of the \textsc{ec} gadget. Case (ii) takes an incoming state without any error (see Eq.~\eqref{eq:entangled_state}) to the state $|11\rangle |\phi\rangle \otimes (\ket{11}\ket{\phi})_f$ (see Eq.~(8) of the main text). This is detected by giving an odd parity, $s_1=1$, and subsequently, $u_1=1$ and $v_1=1$, and after being disentangled with the flag qubits, will be corrected by the now no-fault recovery unit (since the single allowed fault in the \textsc{ec} gadget occurred in this parity measurement); see Table~\ref{tab:updatedsyn}.
	\item A faulty \textsc{cnot} at location $2$ involves again two cases: (i) fault at control; (ii) fault at target. Case (i) again is not detected in this parity measurement, and will present as an outgoing damping error. Case (ii) also results in no error as the ancilla state is in the state $|0\rangle$ right after the \textsc{cnot}, a state immune to the effects of $\cF$; $s_1$ remains as $0$ in this case.
	\item A faulty $Z$ measurement on the ancillas introduces no errors to the syndrome extraction -- the ancilla qubits, assuming no incoming errors, remain in the state $|0\rangle$, immune to the damping error.
\end{itemize}

We now consider the effect of an undetected error due to a fault in the syndrome extraction unit, which is denoted as $B_i$ in Tab.~\ref{tab:R0syn}. For concreteness, consider an error on the first data qubit, errors on the other qubits can be understood in the same manner. It can be easily checked that the state after the damping error, $a\ket{0111}\ket{1111}_f + b\ket{0100}\ket{1100}_f$, becomes the following state after passing through the $\cM_1$ unit and the decoupling step
\begin{equation*}
	\begin{split}
		(\tfrac{a}{\sqrt{2}}(\ket{0111} \pm \ket{1000}) + \tfrac{b}{\sqrt{2}}(\ket{0100} \pm \ket{1011}))\ket{1000}_f
	\end{split}
\end{equation*} 
where plus or minus sign depends on the outcome of the measurement in the $\cM_1$ unit. We can see that the first flag qubit is flipped and there is an X or Y error on the first data qubit. In either case, from the flag syndrome and the follow-up parity measurement, we know that the first data qubit is damped and can correct correspondingly. Note that the syndrome extraction alone cannot distinguish between X errors on data qubit 1 and 2, therefore, the flag qubits are necessary in this case to make the \textsc{ec} gadget fault tolerant.

We next move on to the $\cM_1$ unit that appears in Fig.~1 and is detailed in Fig.~2(f) of the main text. A fault at the target of any \textsc{cnot} coupled to a data qubit, denoted as $A_i$ in Tab.~\ref{tab:R0syn}, causes a damping error in the data qubit connected to it. This error is detected by the flag syndrome bits and corrected by the recovery $\cR$ since it flips the flag qubit coupled to the damped data qubit. On the other hand, a fault at the target of any \textsc{cnot} coupled to a flag qubit, denoted as $A_i^{(f)}$ in Tab.~\ref{tab:R0syn}, also flips the flag qubit, but propagates $(I+Z)$ error to the data qubits. By a parity measurement at the end, we are able to distinguish this with the previous case, therefore, correctly recover the encoded state. What is more complicated is a fault at the ancilla qubit. We consider the possible faults at different locations in the $\cM_1$ unit below, and explain how they are mitigated.
\begin{itemize}
	\item A fault in the preparation of the $\ket{+}$ state, denoted as $C_0$, causes the initial state of the ancilla qubit to be $\ket{0}$. This means that the $\cM_1$ unit has no effect on the data qubits and the flag qubits are left in $\ket{0000}$ state after being disentangled with the data qubits. The $X$ measurement at the end will give random outcome but we never use this outcome to decode anything. 
	\item A fault at the control of the first \textsc{cnot}, denoted as $C_1$, propagates $X$ errors to data qubits 2, 3, 4, and to all flag qubits, which is equivalent to an $X$ error on data qubit 1, up to a stabilizer. These $X$ errors on data qubit 2, 3, and 4 in turns propagate through the \textsc{cnot}s after the $\cM_1$ unit and flip flag qubits 2, 3, and 4. The overall effect is that data qubit 1 and flag qubit 1 are flipped, hence, the flag syndrome is $(1000)$ and we are able to correctly apply a bit flip to the first data qubit. 
	\item A fault at the control of the second \textsc{cnot}, denoted as $C_2$, propagates $X$ errors to data qubit 3, 4, and to all flag qubits, which is equivalent to a logical $X$ error. In this case, the flag qubits give an unique syndrome $(1100)$ which has two bit flips instead of single bit flip as in all the other cases. 
	\item A fault at the control of the third \textsc{cnot}, denoted as $C_3$, propagates $X$ errors to data qubit 4, and to all flag qubits. After the decoupling step, the flag syndrome is $(1110)$, the recovery unit will be able to recognize and correct it.
	\item A fault at the control of the fourth \textsc{cnot}, denoted as $C_4$, does not cause any error to the data qubit, but flips all the flag qubit. Therefore, the flag syndrome is $(1111)$ and we conclude that no data qubit is damped.   
	\item A fault at the control of the fifth \textsc{cnot}, denoted as $C_5$, propagates $X$ errors to flag qubit 2, 3, and 4, hence, makes the flag syndrome $(0111)$. This is a unique syndrome and we conclude that there is no data error in this case. 
	\item A fault at the control of the sixth \textsc{cnot}, denoted as $C_6$, propagates $X$ errors to flag qubit 3 and 4, hence, makes an unique flag syndrome $(0011)$ and we also conclude that there is no error in the data qubits. 
	\item A fault at the control of the seventh \textsc{cnot}, denoted as $C_7$, propagates an $X$ error to flag qubit 4, hence, the flag syndrome $(0001)$. A parity measurement after that is necessary to distinguish this with $A_1^{(f)}$ or $D_1$ fault.
	\item A fault at the control of the eighth \textsc{cnot}, denoted as $C_8$, or a fault at the $X$-measurement does not propagate any error to data and flag qubits. It may change the measurement outcome, which has no consequence for us.
	
\end{itemize}

Finally, a fault at the control of the \textsc{cnot}s used to disentangle data and flag qubits causes a damping error that is undetected by the current \textsc{ec} gadget, appeared as an error in the outgoing state, to be dealt with by the next \textsc{ec} gadget. Meanwhile, the target of those \textsc{cnot}s or $Z$-measurements cannot be damped since the flag qubits are in state $\ket{0000}$ at this step.

Tab.~\ref{tab:R0syn} summarizes all the fault locations discussed above with the corresponding error on the data qubits and syndrome. The column denoted as "$\cP_{12}$ or $\cP_{34}$" indicates which parity measurement should be performed in the second syndrome extraction unit, the other parity measurement is not necessary and always gives a trivial outcome. We note that the syndrome is not unique for each fault location, but it is enough to determine the damped qubit, thereby, enough to correct $E_1$, $E_1^{\dagger}$, $X$, or $Z$ error on that qubit.

This covers all possibilities for a single fault in the \textsc{ec} gadget, giving rise either to no more than one damping error to the outgoing state, i.e., (P2) holds.

\begin{table}[h] 
	\footnotesize
	\begin{center}
		\begin{tabular}{c | c | c | c | c | c || l}
			$s_1$ \ & \ $s_2$ \ & \ $u_1$ \ & \ $v_1$ \ & \ $u_2$ \ & \ $v_2$ \ & \textbf{Diagnosis}\\ 
			\hline\hline
			0&0&$\times$&$\times$&$\times$&$\times$& no damping error or undetected fault\\
			\hline
			1&0&0&1&$\times$&$\times$& Qubit 1 is damped \\
			\hline
			1&0&1&0&$\times$&$\times$& Qubit 2 is damped \\ 
			\hline
			1&0&1&1&$\times$&$\times$& Fault in \textsc{cnot} at location 1 of $\cP_1$ \\ 
			\hline
			0&1&$\times$&$\times$&0&1& Qubit 3 is damped \\ 
			\hline
			0&1&$\times$&$\times$&1&0& Qubit 4 is damped \\ 
			\hline
			0&1&$\times$&$\times$&1&1& Fault in \textsc{CNOT} at location 1 of $\cP_2$ \\ 
		\end{tabular}
	\end{center}
	\caption{\label{tab:updatedsyn} Updated version of Table~\rom{1} of the main text, with added syndromes and diagnoses for faults in the \textsc{ec} gadget.}
\end{table}

\begin{table}
	\begin{center}
		\begin{tabular}{|c|c|c|c|c|}
			\hline
			Fault & Error & $\textbf{r}$ & $\cP_{12}$ or $\cP_{34}$ & $\cR$ \\
			\hline
			None & $I$ & (0000) & None & None \\
			\hline
			$A_1$ & $E_1$ & (1000) & $\cP_{12} \rightarrow 1$ & $X_1/Z_1$ \\
			$A_2$ & $E_2$ & (0100) & $\cP_{12} \rightarrow 1$ & $X_2/Z_2$ \\
			$A_3$ & $E_3$ & (0010) & $\cP_{34} \rightarrow 1$ & $X_3/Z_3$\\
			$A_4$ & $E_4$ & (0001) & $\cP_{34} \rightarrow 1$ & $X_4/Z_4$\\
			\hline
			$A_1^{(f)}$ & $I+Z_1$ & (1000) & $\cP_{12} \rightarrow 0$ & $Z_1$ \\
			$A_2^{(f)}$ & $I+Z_2$ & (0100) & $\cP_{12} \rightarrow 0$ & $Z_2$ \\
			$A_3^{(f)}$ & $I+Z_3$ & (0010) & $\cP_{34} \rightarrow 0$ & $Z_3$ \\
			$A_4^{(f)}$ & $I+Z_4$ & (0001) & $\cP_{34} \rightarrow 0$ & $Z_4$ \\
			\hline
			$B_1$ & $X_1$ or $Z_1X_1$ & (1000) & $\cP_{12} \rightarrow 1$ & $X_1/Z_1$ \\
			$B_2$ & $X_2$ or $Z_2X_2$ & (0100) & $\cP_{12} \rightarrow 1$ & $X_2/Z_2$ \\
			$B_3$ & $X_3$ or $Z_3X_3$ & (0010) & $\cP_{34} \rightarrow 1$ & $X_3/Z_3$ \\
			$B_4$ & $X_4$ or $Z_4X_4$ & (0001) & $\cP_{34} \rightarrow 1$ & $X_4/Z_4$ \\
			\hline
			$C_0$ & $I$ & (0000) & None & None \\
			$C_1$ & $X_1$ & (1000) & $\cP_{12} \rightarrow 1$ & $X_1$ \\
			$C_2$ & $X_3X_4$ & (1100) & None & $X_3X_4$ \\
			$C_3$ & $X_4$ & (1110) & None & $X_4$ \\
			$C_4$ & $I$ & (1111) & None & None \\
			$C_5$ & $I$ & (0111) & None & None \\
			$C_6$ & $I$ & (0011) & None & None \\
			$C_7$ & $I$ & (0001) & $\cP_{34} \rightarrow 0$ & $I$ \\
			$C_8$ & $I$ & (0000) & None & None \\
			\hline
			$D_1$ & $I$ or $Z_1$ & (1000) & $\cP_{12} \rightarrow 0$ & $Z_1$ \\
			$D_2$ & $I$ or $Z_2$ & (0100) & $\cP_{12} \rightarrow 0$ & $Z_2$ \\
			$D_3$ & $I$ or $Z_3$ & (0010) & $\cP_{34} \rightarrow 0$ & $Z_3$ \\
			$D_4$ & $I$ or $Z_4$ & (0001) & $\cP_{34} \rightarrow 0$ & $Z_4$ \\
			\hline
		\end{tabular}
	\end{center}
	\caption{\label{tab:R0syn} Summary of fault locations and the corresponding error and syndrome for \textsc{ec} gadget in Fig.~1 of the main text. The first column denotes all the fault locations mentioned in the proof. The second column is the errors on the data qubits at the point right after measuring all the flag qubits. The third column is the corresponding flag syndromes. The fourth column indicates which parity measurement should be performed in the second syndrome extraction unit in Fig.~1(b) of the main text and the expected outcome. $\cP_{12}$ and $\cP_{34}$ mean the parity measurements on data qubits 1\&2 and 3\&4, respectively. The last column gives the corresponding recovery gate. Note that a $Z$ correction is applied only if the outcome of the measurement of $XXXX$ in the $\cR$ unit is $+1$. }
\end{table}

\subsection{Logical $X$ gadget}\label{app:logicalx}
Next, we demonstrate how the logical $X$ gadget described in Fig.~1 of the main text is tolerant against single faults, as stated in property (P5) in SM Sec.~\ref{app:faulttolerance}.

First, we note that an $\cF_z$ error passes through the logical $X$ in the same way as in the $\textsc{ec}$ gadget because $\cF_z$ (anti)commutes with a physical $X$ gate. Therefore, it will be killed off by the $\cM_1$ unit or result in an $\cF_z$ in the outgoing state, as explained in SM Sec.~\ref{app:ECFT}.

\begin{itemize}
	\item An error in the incoming state or a fault in the entangling step between the data qubits and the flag qubits is detected by the following syndrome extraction unit $\cS$ (see Fig.~1 of the main text). The circuit after that can be assumed to have no fault, hence, recover for the damping error and apply the transversal $\overline{X}=XXII$ correctly.
	\item An error before one of the physical $X$ gates (due to a fault in the syndrome extraction unit) conjugates through the $X$ gate as $E_1^{\dagger}= (1-Z)X$, which in turn becomes an $E_1$ or an $E_1^{\dagger}$ after the $\cM_1$ unit. The flag syndromes are swapped as compared to the case of $B_i$ faults in Tab.~\ref{tab:R0syn}, namely, $(r_1r_2r_3r_4) = (0111)$ for an $E_1$ error and $(r_1r_2r_3r_4) = (1000)$ for an $E_1^{\dagger}$ error. In either case, a bit flip is correctly applied to the first data qubit. 
	\item A fault at other locations is already covered in the analysis of the \textsc{ec} gadget, see SM Sec.~\ref{app:ECFT}.  
\end{itemize}	

Thus the logical $X$ gadget is tolerant up to single-qubit errors, thereby satisfying the desired fault tolerance properties.

\subsection{Logical \textsc{cz} gadget}\label{app:cz}

Finally, we show that the \textsc{cz} gadget in Fig.~3 of the main text satisfies the fault tolerance property (P5) stated in SM Sec.~\ref{app:faulttolerance}.

First, note that an $\cF_z$ in the incoming state or due to any faulty \textsc{cz}s passes through the gadget and is killed off by the \textsc{ec} gadget at the end. A $\cF_z$ due to a faulty component in the \textsc{ec} gadgets is either killed off by the \textsc{ec} gadgets themselves or causes at most one $\cF_z$ at the outgoing state, as explained in Sec.~\ref{app:ECFT}.

\begin{figure}[h]
	\centering
	\includegraphics[scale=.4]{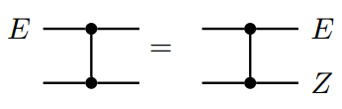}
	\caption{A damping error $E$ before a \textsc{cz} gate. }
	\label{fig:cphase_analysis}
\end{figure}

Next, consider the case of a damping error propagating through a two-qubit \textsc{cz} gate, as shown in Fig.~\ref{fig:cphase_analysis}.

An incoming damping error right before the control (target), of a physical \textsc{cz} gate, in Fig.~3 of the main text, propagates as a damping error $E$ at the control (target) and a phase error $Z$ at the target (control). This is detected in the $\cM_1$ unit of the target (control) block via a nontrivial outcome, with $c_1=1$. It is then corrected by applying a $Z$ operator to the appropriate qubit in the control (target) block (see Tab.\ref{tab:CZsyn}). Contrastingly, a damping error after the control (target) of a physical \textsc{cz} gate, in Fig.~3 of the main text leads to a damping error at the output of the same block, without propagating a Z error in the target (control) of the other block. This scenario is captured by a trivial outcome, with $c_1=0$, in the $\cM_1$ unit of the target (control) block.
Finally, a fault anywhere in the control or target block leads to at most one error in the outgoing state of one of the two blocks, as shown in SM Sec.~\ref{app:ECFT}. Therefore, we conclude that our transversal \textsc{cz} gadget satisfies the desired fault tolerance properties. 

\begin{table}
	\begin{center}
		\begin{tabular}{|c|c|}
			\hline
			Damped qubit in one block & Propagated error to the other block \\
			\hline
			Qubit 1 & $Z_1$ \\
			Qubit 2 & $Z_3$ \\
			Qubit 3 & $Z_2$ \\
			Qubit 4 & $Z_4$ \\
			\hline
		\end{tabular}
	\end{center}
	\caption{\label{tab:CZsyn} Summary of $Z$ errors propagated to one block due to an $\cF_a$ error before $\overline{\textsc{CZ}}$ in the other block. In all these cases, the $\cM_1$ unit in the trailing \textsc{EC} of the no-damping block gives outcome $+1.$}
\end{table}

\subsection{Logical $\textsc{ccz}$ gadget}\label{app:ccz}
\begin{figure}[h]
	\centering
	\includegraphics[width=0.9\columnwidth]{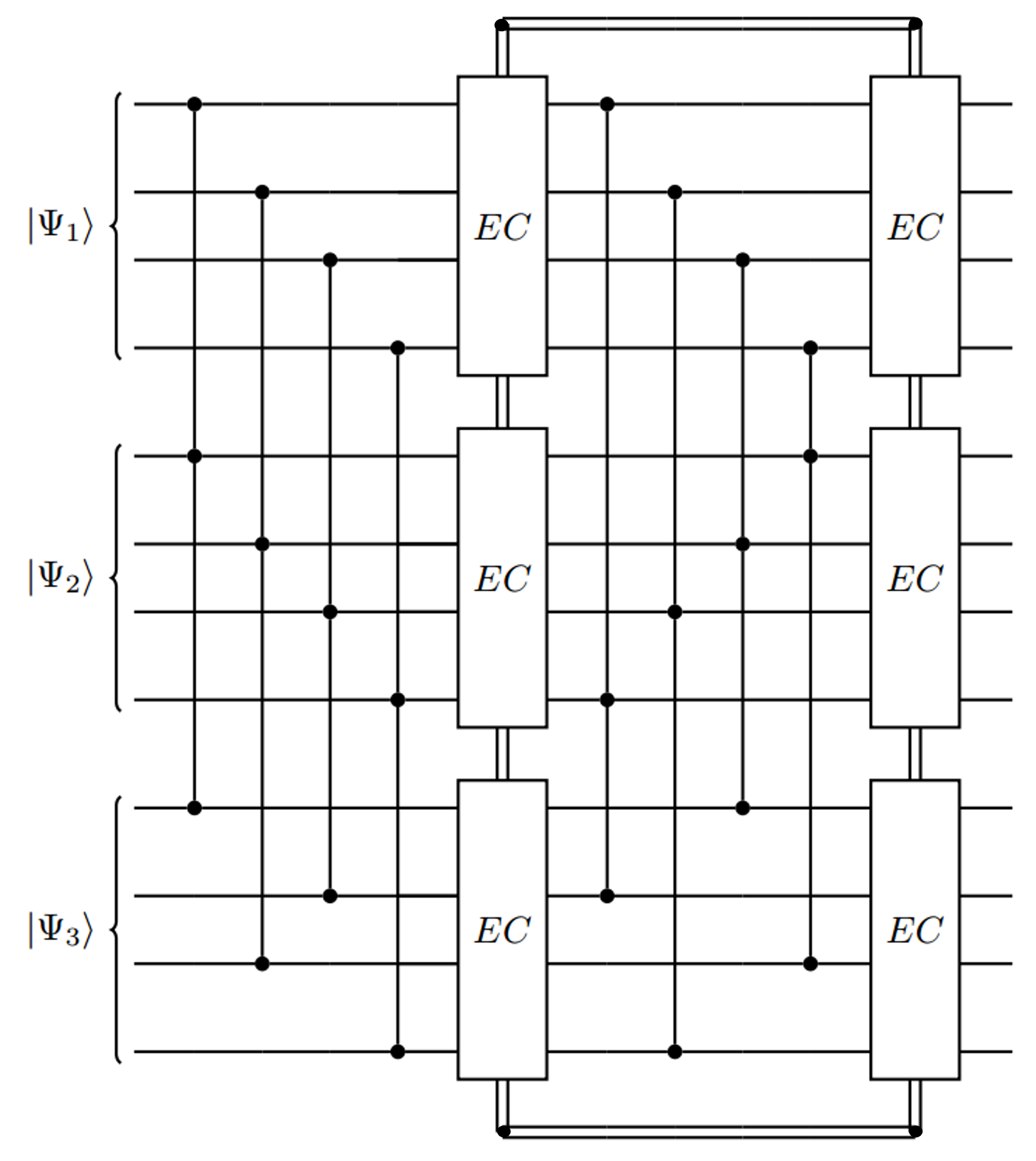}
	\caption{Transversal fault-tolerant logical $\textsc{ccz}$ gadget. Similar to the logical $\textsc{cz}$ gadget, syndrome bits from different $\textsc{ec}$ gadgets are combined to catch propagated $Z$ errors.} 
	\label{fig:ccz}
\end{figure}

The transversal logical $\textsc{ccz}$ is shown in Fig.~\ref{fig:ccz}. 

\section{Preparation gadgets}\label{sec:logical_prep}

Here, we explain how to construct the preparation gadgets for the states $|0\rangle_L$ and $|+\rangle_L$, which form part of the universal set of logical gadgets discussed in Sec.~\rom{5} of the main text. In addition, the fault-tolerant preparations of the resource states $|\Phi_S\rangle$ and $|\Phi_T\rangle$, used in the $S$ and $T$ gadgets of Sec.~\rom{5}, are also given here.

\subsection{Logical states}

Armed with the ability to prepare the Bell state fault-tolerantly, as shown in SM Sec.~\ref{sec:prep}, we can obtain fault-tolerant preparations of the logical states $|0\rangle_L$ and $|+\rangle_L$. The $|+\rangle_L$ state is straightforward: $|+\rangle_L$ is simply two copies of $|\beta_{00}\rangle$, that is, 
\[ |+\rangle_L=|\beta_{00}\rangle\otimes|\beta_{00}\rangle.\]
Here, the preparation unit for the $\ket{+}_L$ state consists of two preparations of the $\ket{\beta_{00}}$ state. A fault-tolerant preparation requires that if at most one fault occurs in this combined circuit, the output has at most one error. This is different from, say, the Hadamard circuit discussed below which consists of $\overline{\textsc{cz}}$, $\ket{+}_L$ and $\overline{X}$ units, where one fault is allowed in each of those components. Here, even though we say that the preparation of $\ket{\beta_{00}}$ is fault tolerant, we don't allow one fault in each preparation of $\ket{\beta_{00}}$ when talking about preparation of $\ket{+}_L$ because there is no \textsc{ec} gadget attached to the preparation of $\ket{\beta_{00}}$ and one fault in each preparation of $\ket{\beta_{00}}$ may result in two errors in the outgoing $\ket{+}_L$ state.

\begin{figure}[H]
	\begin{center}
		\includegraphics[width=0.6\columnwidth]{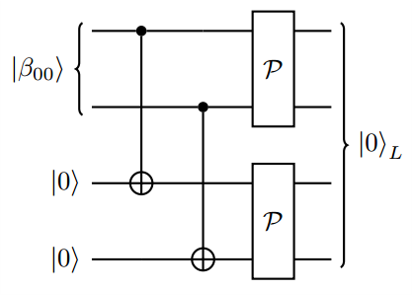}
		\caption{Fault-tolerant preparation of $|0\rangle_L$. The input Bell state $|\beta_{00}\rangle$ is assumed to have been prepared fault-tolerantly by the preparation circuit of Fig.~\ref{fig:BellPrep}.}
		\label{fig:zero}
	\end{center}
\end{figure}

To get the state $|0\rangle_L$, we start with a single copy of a fault-tolerantly prepared $|\beta_{00}\rangle$ and make use of the circuit in Fig.~\ref{fig:zero}, with two additional ancillas initialized to $|0\rangle$. The prepared state is accepted only when both the parity measurements are even. The fact that the preparation circuit is fault-tolerant can be seen as follows. The off-diagonal error $\cF_z$ of a single fault in the circuit passes through the parity measurements and causes only one $\cF_z$ error in the outgoing state. Meanwhile, the damping error $\cF_a$ in the preparation of $\ket{\beta_{00}}$ or in two \textsc{cnot}s is detected by the parity measurements, and thereby, rejected. An undetected fault in the parity measurements leads to only one error in the outgoing state. In any case, the state is either rejected, or accepted with at most one error.

\subsection{Resources states}

\begin{figure}[H]
	\centering
	\includegraphics[width=0.7\columnwidth]{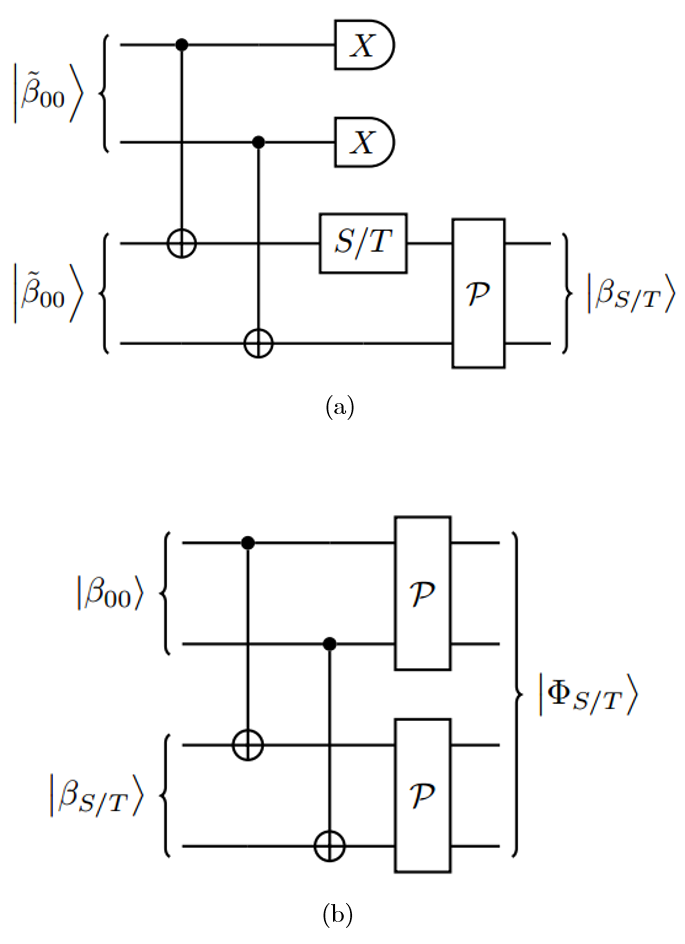}
	\caption{(a) Fault-tolerant preparation of $|\beta_{S/T}\rangle$ using two copies of the faulty Bell state $\ket{\tilde{\beta}_{00}}$. (b) Fault-tolerant preparation of $|\Phi_{S/T}\rangle$ using states $|\beta_{00}\rangle$ and $|\beta_{S/T}\rangle$ which are assumed to be fault-tolerantly prepared.} 
	\label{fig:resource}
\end{figure}

Next, we demonstrate fault-tolerant preparation gadgets for the two-qubit states $\ket{\Phi_{S}}$ and  $\ket{\Phi_{T}}$, which act as resource states for constructing the logical $S$ and $T$ gates, respectively:  
\begin{align}\label{eq:resource}
	\ket{\Phi_S} &= \frac{1}{\sqrt{2}}{\left(\ket{0}_L + \mi\ket{1}_L\right)} \\
	\textrm{and }\quad    \ket{\Phi_T} &= \frac{1}{\sqrt{2}}{\left(\ket{0}_L + \mathrm{e}^{\mi\pi/4}\ket{1}_L\right)}.\nonumber
\end{align}
The resource states $\ket{\Phi_S}$ and $\ket{\Phi_T}$ can be prepared and verified as shown in Fig.~\ref{fig:resource}, starting with a fault-tolerant preparation of the states $|\beta_{S}\rangle$ and $|\beta_{T}\rangle$, which are local-unitary equivalents of $|\beta_{00}\rangle$:
\begin{align}
	|\beta_{S}\rangle &\equiv \frac{1}{\sqrt 2}{\left(|00\rangle+ \mi |11\rangle\right)} \nonumber \\
	\textrm{and}\quad |\beta_{T}\rangle &\equiv \frac{1}{\sqrt 2}{\left(|00\rangle+ \mathrm{e}^{\mi \pi/4} |11\rangle\right)}.
\end{align} 
Fault-tolerant preparation units for the states $|\beta_{S/T}\rangle$ are shown in Fig.~\ref{fig:resource}(a), using a circuit similar to that for the preparation of $|\beta_{00}\rangle$ (Fig.~\ref{fig:BellPrep}). In each case, we accept the output state only when the $X$-measurement outcomes are of even parity and the parity measurement provides a trivial outcome. Using $|\beta_{S/T}\rangle$ states, we then prepare and verify the resource states $|\Phi_{S/T}\rangle$ in Eq.~\eqref{eq:resource} as shown in Fig.~\ref{fig:resource}(b). We accept the prepared state only when both the parity measurements show trivial outcomes. The fault-tolerant property of the preparation of $\ket{\Phi_{S/T}}$ can be understood in a similar manner to the preparation of the $\ket{0}_L$ state. A single fault in the circuit in Fig.~\ref{fig:resource}(b) leads to the state being rejected or accepted with at most one error.

\section{Pseudothreshold calculations}\label{app:thresh_calc}

We describe here the technical details of our pseudothreshold calculation, for the memory gadget and the extended \textsc{cz} gadget and its connection with the infidelity between the noisy and the ideal output. We assume that the inputs to the gadgets do not have any errors and explicitly count the total number of malignant faults of $O(p^2)$ which will cause a given gadget to fail. $O(p^2)$ faults include phase fault at a single position and damping faults at two different positions.

\subsection{Relating fault counts to encoded infidelity}\label{app:infidelity}
In Secs.~VIA and B of the main text, we exhaustively counted the number of malignant fault pairs and all fault triples that can lead to an erroneous output state even after the decoder. This has to be connected with an actual figure-of-merit---the infidelity in our case---for the correctness of the output state. Here, we provide the proof of that connection as stated in Sec.~VI of the main text [leading to Eq.~(10) there].

Recall that, to operationally gauge whether the output state is correct or not, we pass the output of an extended gadget through an ideal decoder to strip off the correctable $O(p)$ terms. The fault tolerance of the EC unit guarantees that no $O(p)$ terms give an error after the ideal decoder. The final state $\rho$ from the output of an ideal decoder thus obtained can then be expanded as
\begin{equation}\label{eq:output}
	\rho = c_0\rho_0 +  p^2\sum_{j}\epsilon_{j} + O(p^3).
\end{equation}
Here, $\rho_0$ is the ideal output state, occuring with weight $c_0$. The summation over $j$ explicitly enumerates all malignant two-fault paths, which contribute second-order [$O(p^2)$] terms. All the correctable, i.e., benign, second-order terms have already been absorbed into the coefficient $c_0$. The $\epsilon_j$s are general operators that need not be states, i.e., not necessarily positive semi-definite nor of trace 1.

We enumerate the different cases of order $p^2$ that can arise from a fault at a single location or faults at a pair of locations:\label{para:fault}
\begin{itemize}
	\item Recall from Eq.~(4) of the main text that a $Z$ error of the form $Z(\cdot)Z$ is of $O(p^2)$. A $Z$ error could lead to an output with either no error or a $Z_L$ error. A $Z_L$ error can occur since the states after a single $Z$ error on each qubit become indistinguishable, therefore, one could leave a relative phase while trying to correct for it.
	\item A fault pair at locations denoted as A and B has the form $\cF^{\otimes 2} = (\tfrac{1}{2}\cF^{(A)}_z + \cF^{(A)}_a)\otimes(\tfrac{1}{2}\cF^{(B)}_z + \cF^{(B)}_a)$. From the definition of a fault in Eq.~(4) of the main text, we can see that a fault pair leads to the following terms (up to constant factors): 
	\begin{itemize}
		\item two $\cF_z$s: gives $Z^{(A)}Z^{(B)}(\cdot) + (\cdot)Z^{(B)}Z^{(A)}$ and $Z^{(A)}(\cdot)Z^{(B)} + Z^{(B)}(\cdot)Z^{(A)}$, which would end up as the identity or operator of the form  $Z_L(\cdot) + (\cdot)Z_L$ which are generally not positive semi-definite. Note that such a term will survive at the decoder output and does not get removed by the $XXXX$ measurement.
		\item one $\cF_z$ and one $\cF_a$: gives $E^{(A)}Z^{(B)}(\cdot)E^{\dagger(A)} + E^{(A)}(\cdot)Z^{(B)}E^\dagger{(A)}$ and another term with the roles of A and B swapped. An  $\cF_a$ error can be thought of as an $X$ error once the superposition is restored in the code space. Therefore, failing to correct for $\cF_a$ would lead to $X_L$. An $\cF_Z$ followed by an $\cF_a$ could lead to applying a relative phase on the state with a damping error. Trying to correct for this combination of errors could lead to at most an $X_L$ or $Z_L(\cdot) + (\cdot)Z_L$ error on the output state.
		\item two $\cF_a$s: gives $E^{(A)}E^{(B)}(\cdot)E^{\dagger(B)}E^{\dagger(A)}$. Performing the $XXXX$ measurement in the recovery rebuilds the loss of superposition due to damping errors. Therefore, trying to correct for two damping errors is like trying fix for two $X$ errors which can lead to an output state with an $X_L$ or $(1+Z_L)$ error.
	\end{itemize}
\end{itemize}

Considering these possibilities, we see that the operators $\epsilon_j$ in Eq.~\eqref{eq:output} can be written explicitly as
\begin{equation}\label{eq:rhoj}
	\epsilon_j = c_j \rho_j + \tfrac{d_j}{2}(Z_L\rho_0 + \rho_0Z_L),
\end{equation}
for some state $\rho_j$ in the code space and some coefficients $c_j$, $d_j$ $\in$ $\mathbb{R}$. Note that the weights $c_j$s can be verified, by following the error correction procedure, to be positive constants.


The ideal output state $\rho_0$ in Eq.~\eqref{eq:output} from an extended unit is a pure state in the code space, which we can write as $\rho_0\equiv |\Psi\rangle\langle\Psi|$. The infidelity of the output state $\rho$ in Eq.~\eqref{eq:output} with respect to the ideal output state $\ket{\Psi}$ is then,
\begin{equation}
	\mathrm{IF}(\rho, \ket{\Psi}\bra{\Psi}) = 1 - \bra{\Psi}\rho\ket{\Psi} = \Tr(\rho) - \bra{\Psi}\rho\ket{\Psi},
\end{equation}
where
\begin{align*}
	\mathrm{Tr}(\rho) 
	&= c_0 + p^2\sum_j \left(c_j + d_j\bra{\Psi}Z_L\ket{\Psi}\right) + O(p^3),\\
	\langle\Psi|\rho|\Psi\rangle
	&= c_0 + p^2 \sum_j \left(c_j\langle\Psi|\rho_j|\Psi\rangle + d_j\langle\Psi|Z_L|\Psi\rangle\right) + O(p^3).
\end{align*}
Hence, we have,
\begin{align}
	\mathrm{IF}(\rho, \ket{\Psi}\bra{\Psi}) &= p^2\sum_{j}c_j(1-\bra{\Psi}\rho_j\ket{\Psi}) +O(p^3) \nonumber\\
	& \leq p^2\sum_j c_j  + O(p^3)\label{eq:infide}
\end{align}
where, in the last line, we have used the fact that $\bra{\Psi}\rho_j\ket{\Psi} \geq 0$, true for any state $\rho_j$. The infidelity is then bounded as
\begin{equation}
	\mathrm{IF}(\rho, \ket{\Psi}\bra{\Psi})\leq Cp^2+Bp^3, 
\end{equation}
where $C$ is the total number of second-order faults or malignant fault pairs (as counted in Secs.~VIA and B of the main text), and $B$ is the number of all possible ways that the gadget can have a third-order fault. Recall from Sec.~VI of the main text that when counting $C$, we also took care of additional multiplicative factors, which correspond exactly to the $c_j$ coefficients in Eq.~\eqref{eq:infide}.

\subsection{Memory pseudothreshold}\label{app:memory}

We now calculate the pseudothreshold for the memory gadget in Fig.~5 of the main text. We first count the malignant fault pairs due to two damping faults (assuming a no-error input to the memory gadget) leading to an output that is uncorrectable. This could happen in one of three possible ways: (1) two faults within the \textsc{ec} gadgets, (2) two faults within the rest locations, or, (3) one fault in the \textsc{ec} gadget and one fault in the rest locations.

\begin{enumerate}
	\item {\bf Malignant pairs within  an \textsc{ec} gadget:} We count the total number of malignant pairs within an \textsc{ec} gadget, shown in Fig.~1 of the main text. Depending on the outcomes of the syndrome extraction unit $\cS$, the \textsc{ec} can take different paths. For easy counting, we further divide the \textsc{ec} unit into smaller parts, numbered as follows.
	\begin{enumerate}
		\item[1.] Entangling step between data and flag qubits, including the first 4 \textsc{cnot}s.
		\item[2.] Syndrome extraction unit $\cS$.
		\item[3.] Disentangling step between the data and flag qubits in case a damping is detected. 
		\item[4.] Recovery unit $\cR$ in case a damping is detect.
		\item[5.] $\cM_1$ unit.
		\item[6.] Disentangling step between the data and flag qubits in case no damping is detected.
		\item[7.] Parity measurements $\cP_{12}$ and $\cP_{34}$.
		\item[8.] Recovery unit $\cR$ in case no damping is detected.
	\end{enumerate}
	The matrix below represents the number of malignant pairs with each fault in the corresponding parts.
	\[
	\begin{blockarray}{cccccccccc}
		& & 1 & 2 & 3 & 4 & 5 & 6 & 7 & 8\\
		\begin{block}{cc(cccccccc)}
			1 & & 112 \\
			2 & & 263 & 61 \\
			3 & & 18 & 4 & 0 \\
			4 & & 50 & 17 & 0 & 0 \\
			5 & & 650 & 617 & 0 & 0 & 1767 \\
			6 & & 56 & 75 & 0 & 0 & 430 & 25 \\
			7 & & 96 & 70 & 0 & 0 & 618 & 0 & 0 \\
			8 & & 92 & 58 & 0 & 0 & 463 & 0 & 0 & 0 \\
		\end{block}	
	\end{blockarray}
	\]
	Each entry in the matrix is the number of malignant pairs with two faults in the parts corresponding to row and column label. It is counted by choosing any one position in each part, inserting damping faults and checking if the final state is correctable or not. In total, there are $5542$ malignant pairs in one \textsc{ec} gadget. 
	
	\item {\bf Malignant pairs in the resting locations:} There are $4$ locations when the qubits rest or there are $4$ applications of Identity gates leading to $C(4,2)=6$ malignant pairs.
	
	\item {\bf 1 fault in the leading \textsc{ec} and $1$ fault in the resting locations:}
	Most of the single faults inside an \textsc{ec} is detected and corrected by the \textsc{ec} itself. However, there are single faults that lead to a single error at the outgoing state of the \textsc{ec}. They include faults at the control of one of four \textsc{cnot}s at the disentangling step, which cause $\cF_a$ or $\cF_z$ errors, and faults at the target of one of eight \textsc{cnot}s in $\cM_1$ unit, which cause $\cF_z$ errors. 
	This error in turn can combine of one error at one of four rest location, leading to an uncorrectable error.  In total, we have $28$ such malignant pairs.
	
	\item{\bf 1 fault in the trailing \textsc{ec} and $1$ fault in the rest location:}
	A fault in one of four rest location can combine with a fault in the trailing \textsc{ec} to cause an uncorrectable error. In total, there are $224$ pairs. In case the error in the rest location is $\cF_a$, $\cM'$ in Fig.~1 of the main text will be the Identity. Otherwise, if it is a $\cF_z$ error, $\cM'$ will be $\cM_1$ unit. 
	
	\item{\bf 1 fault in each \textsc{EC}: }
	Faults at the control of \textsc{cnot}s in part 6 (disentangling step) of the leading \textsc{ec} can combine with one fault in the trailing \textsc{ec} in the same way as faults in rest location. There are $448$ malignant pairs due to this. In addition, $\cF_z$ error due to faults at the target of \textsc{cnot}s in part 5 ($\cM_1$ unit) of the leading \textsc{ec} also can combine with faults in the trailing \textsc{ec}. There are $254$ malignant pairs due to this. Therefore, in total, there are $702$ malignant pairs for this case.   
\end{enumerate}

Finally, we count the malignant faults leading to $Z$ errors in the memory gadget in Fig.~5 of the main text. For an \textsc{ec} gadget, this includes $11$ positions in part 1, $6$ positions in part 2, $36$ positions in part 5, and $4$ positions in part 6. Hence, 57 positions for an \textsc{ec} gadget. For the whole memory gadget, there are $29$ positions in total, taken into account the multiplication factor of $\tfrac{1}{4}$.

Therefore, the total number of malignant pairs due to damping faults and malignant positions due to $Z$ errors is given by, $C = 6531$. Furthermore, there are at most $181$ locations in one \textsc{ec} gadget, therefore $B= {366\choose 3} + {366\choose 2} + 366 = 8,171,621$.

\subsection{Computational pseudothreshold}\label{app:cphase_threshold}

The computational pseudothreshold, as explained in the main article, is determined by the extended \textsc{cz} gadget. The counting for the extended \textsc{cz} can be done in very similar manner as for the memory gadget. We show here again the matrix whose rows and columns correspond to each part in Fig.~6 of the main text and whose entries are the total malignant pair contributions from the corresponding parts.

\[
\begin{blockarray}{ccccccc}
	&  & 1 & 2 & 3 & 4 & 5 \\
	\begin{block}{cc(ccccc)}
		1 & & 0 & & & & \\
		2 & & 48 & 0 & & & \\
		3 & & 718 & 328 & 5542 & &  \\
		4 & & 328 & 718 & 24 & 5542 & \\
		5 & & 28 & 28 & 230 & 230 & 12 \\
	\end{block}	
\end{blockarray}
\]

\begin{enumerate}
	\item{\bf Two faults in one \textsc{ec}:}
	We have already counted the number for this case from the last section of memory gadget.
	
	\item{\bf One fault in $\textsc{ec}_1$ and one fault in $\textsc{ec}_3$:}
	This is the same as the case when one fault is in the leading \textsc{ec} and the other fault is in the trailing \textsc{ec} of the memory unit. By symmetry, this is also the number for one fault in $\textsc{ec}_2$ and one fault in $\textsc{ec}_4$. 
	
	\item{\bf One fault in $\textsc{ec}_1$ and one fault in $\textsc{ec}_4$:}
	Most of the faults will be corrected independently. However, if the fault in $\textsc{ec}_1$ causes $\cF_a$ error and a $Z$ error is propagated to the second data block, then a fault in $\textsc{ec}_4$ may miss this $Z$ error. First of all, faults in $\cM_1$ unit at the preparation of $\ket{+}$, $X$-measurement, or the control of the last \textsc{cnot} in $\cM_1$ unit may lead to wrong outcome, hence, $Z$ error. Secondly, a $\cF_a$ error may cause logical error, for example, if a $Z$ error is propagated to the first data qubit of the second block, then a $\cF_a$ error at data qubit 3 or 4 may cause a logical $Z$ error because $Z_1E_3 = Z_1X_3 + \overline{Z}X_3$. In total, there are $328$ malignant pairs for this case. By symmetry, this is also the number for ne fault in $\textsc{ec}_2$ and one fault in $\textsc{ec}_3$.
	
	\item{\bf One fault in $\textsc{ec}_1$ and one fault in $\textsc{ec}_2$:} 
	Note that $\cF_z$ errors are okay since they are corrected independently by $\textsc{ec}_3$ and $\textsc{ec}_4$. Undetected $\cF_a$ must be due to a fault at control of one of four \textsc{cnot}s in the disentangling step. $\cF_a$ at qubits connected by a \textsc{cz} gate are okay because after a damping, a $Z$ error has no effect. However, $\cF_a$ at qubits not connected by any \textsc{cz} gate lead to a logical $Z$ error in one or both data blocks. Therefore, there are $48$ pairs. 
	
	\item{\bf One fault in $\textsc{ec}_3$ and one fault in $\textsc{ec}_4$: }
	The only case that can fail is when a fault in $\textsc{ec}_3$ leads to non-trivial outcomes of syndrome extraction unit and a fault in $\textsc{ec}_4$ leads to a wrong outcome of $\cM_1$ unit, and vice versa. A fault at the preparation of $\ket{+}$, $X$-measurement and the control of the last \textsc{cnot} in $\cM_1$ unit of $\textsc{ec}_4$ may lead to wrong outcome. There are $8$ positions in $\textsc{ec}_3$ that leads to non-trivial outcomes of syndrome extraction unit: controls in the entangling step and the target of the first two $\textsc{cnot}$s in the syndrome extraction itself. But note that a redundant $Z$ error at data qubit $1$ and $2$ is correctable by an ideal decoder, hence, $4$ among those $8$ locations are safe. Therefore, in total, there are $(4\times3)\times2 = 24$ pairs. 
	
	\item{\bf One fault in \textsc{cz} gadget and one fault in $\textsc{ec}_1$ or $\textsc{ec}_2$:}
	The number is the same as one fault in the leading \textsc{ec} and one fault in the rest location of the memory gadget.
	
	\item{\bf One fault in \textsc{cz} gadget and one fault in $\textsc{ec}_3$ or $\textsc{ec}_4$:}
	The number is the same as one fault in the trailing \textsc{ec} and one fault in the rest location of memory gadget. 
	
	\item{\bf 2 faults in \textsc{cz} gadget:}
	There are 8 positions, hence, maximum $C(8, 2)=28$ pairs. But if two faults are in different block then they are corrected independently. Therefore, we have $12$ pairs left.
	
\end{enumerate}

For a second order $Z$ error, there are $14.25$ positions for one \textsc{ec} unit and $2$ positions for the \textsc{cz} gadget. Therefore we have a total of $14.25 \times 4 + 2 = 59$ malignant positions.  

Therefore, the total number of malignant pairs due to damping errors and malignant fault locations due to $Z$ errors is  given by, $C = 13835$.There are at most $732$ locations in the extended $\textsc{cz}$ gadget, hence $B= {732 \choose 3} + {732 \choose 2} + 732 = 65,371,138$.

\subsection{Pseudothreshold Simulation}\label{app:sim}

In the simulation, to recognize uncorrectable output states which correspond to a failure of the fault-tolerant computation, an ideal decoder that applies the perfect error correction procedure described in Sec.~\rom{3} of the main text is used to strip off the $O(p)$ errors and project the state back into the code space. Specifically, it detects a damping error $\cF_a$ by the syndrome extraction unit. If an $\cF_a$ error is detected, the recovery unit will measure the $XXXX$ stabilizer and apply $X$ or $XZ$ to the damped qubit, corresponding to the measurement outcome. If no $\cF_a$ is detected, the recovery unit will still perform the $XXXX$ measurement to remove the $\cF_z$ error. It might happen that, due to high-order faults, the input to the ideal decoder has a $Z$ error and it is detected by the recovery unit. In this case, the ideal decoder will apply a $Z$ always to the first data qubit, thereby, correctly recovering the state if the $Z$ error is at data qubit $1$ or $2$ but leading to a logical $\overline{Z}$ if the $Z$ error is at data qubit $3$ or $4$.

In our analytical counting of Secs.~VIA and B in the main text, we could easily exclude malignant pairs in the leading EC to avoid double counting. This is, however, difficult implement in the simulation as we apply the full amplitude-damping channel at every step of the circuit. By doing that, we in fact include all malignant pairs in the leading EC as well. Thus, for a fair comparison between the simulation results and our counting, we included all the malignant pairs in the leading EC for the analytical counting when comparing with the simulation. This explains the different values of $\overline{p}_{\rm th}^{(l)}$ in Sec.~VIC, as compared with Sec.~VIA of the main text.

\end{document}